\documentclass{anstrans}

\usepackage{graphicx} 
\usepackage{booktabs} 
\usepackage{microtype} 

\renewcommand{\vec}[1]{\bm{#1}} 

\usepackage{fancyhdr} 

\renewcommand{\thesection}{\Roman{section}.}

\makeatletter
\renewcommand*{\@seccntformat}[1]{\csname the#1\endcsname\hspace{1mm}}
\makeatother

\usepackage{cleveref}
\usepackage{dsfont}
\usepackage{placeins}
\usepackage{color}

\newcommand{\mat}[1]{#1}
\newcommand\diff{\,\mathrm{d}}
\newcommand{\tr}{\mathrm{Tr}}

\newcommand{\sigmaexp}{\vec{\sigma}_\textrm{exp}}
\newcommand{\sigmaexpPart}[1]{\vec{\sigma}_{\textrm{exp},#1}}

\newcommand{\covexp}{\mat{B}}
\newcommand{\covexpBlock}[1]{\mat{B}_{#1}}

\newcommand{\sigmafit}{\vec{\sigma}_\textrm{fit}}
\newcommand{\modpar}{\vec{y}}
\newcommand{\priormodpar}{\vec{y}_0}
\newcommand{\postmodpar}{\vec{y}_1}
\newcommand{\priorcovpar}{\mat{A}_0}
\newcommand{\invpriorcovpar}{\mat{\tilde{A}}_0}
\newcommand{\postcovpar}{\mat{A}_1}
\newcommand{\marcov}{\mat{M}}

\newcommand{\invmarcov}{\mat{\tilde{M}}}
\newcommand{\invmarcovBlock}[1]{\mat{\tilde{M}}_{#1}}

\newcommand{\jacob}{\mat{S}}
\newcommand{\jacobBlock}[1]{\mat{S}_{#1}}

\newcommand{\adjcovexp}{\mat{C}}
\newcommand{\adjcovexpBlock}[1]{\mat{C}_{#1}}
\newcommand{\invadjcovexp}{\mat{\tilde{C}}}
\newcommand{\invadjcovexpBlock}[1]{\mat{\tilde{C}}_{#1}}
\newcommand{\kappavec}{\vec{\kappa}}
\newcommand{\kappavecEl}[1]{\kappa_{#1}}
\newcommand{\lambdavec}{\vec{\lambda}}
\newcommand{\lambdavecEl}[1]{\lambda_{#1}}

\title{Fitting and Analysis Technique for Inconsistent Nuclear Data}
\author{Georg Schnabel}

\institute{
Irfu, CEA, Universit\'e Paris-Saclay, 91191 Gif-sur-Yvette, France
}

\email{georg.schnabel@cea.fr}

\begin{document}
\vspace*{-42pt}
\begin{strip}
\centering{\parbox{153mm}{{\bf Abstract} \itshape - 
Consistent experiment data are crucial to adjust parameters of physics models and to determine best estimates of observables.
However, often experiment data are not consistent due to unrecognized systematic errors.
Standard methods of statistics such as $\chi^2$-fitting cannot deal with this case.
Their predictions become doubtful and associated uncertainties too small.
A human has then to figure out the problem, apply corrections to the data, and repeat the fitting procedure.
This takes time and potentially costs money.
Therefore, a Bayesian method is introduced to fit and analyze inconsistent experiment data.
It automatically detects and resolves inconsistencies.
Furthermore, it allows to extract consistent subsets from the data.
Finally, it provides an overall prediction with associated uncertainties and correlations less prone to the common problem of too small uncertainties.
The method is foreseen to function with a large corpus of data and hence may be used in nuclear databases to deal with inconsistencies in an automated fashion.
}\par}
\vspace*{14pt}
\end{strip}

\section{Introduction}
Evaluated nuclear data help to push forward the development of novel nuclear facilities.
They are needed as input for transport and activation calculations.
Any mistake or inconsistency in the data may distort calculation results and, as a consequence, may lead to suboptimal design choices with regard to efficiency and safety.

The acquisition and determination of consistent nuclear data are non-trivial tasks.
Not all types of nuclear data required for calculations are available from experiments. Especially reaction data at higher energies are scarce.
The remedy is to use predictions of nuclear models or some series expansion and to adjust the parameters to available experimental data.
The adjustment of parameters is usually done by a $\chi^2$-fit or the \textit{Generalized Least Squares} (GLS) method, e.g.~\cite{capote_nuclear_2010}.

The problem with these approaches is the inherent assumption that experiment data are consistent, meaning the data contain what they claim.
Consistent experiment data are an ideal.
Experiments are complex and acquired raw data has to be corrected for many effects such as background noise and detector efficiency.
Already if one such correction has not been done properly, or the associated uncertainties are not estimated well, the experiment data are inconsistent.
The visual signature of inconsistent data sets are points whose error bars mutually exclude each other.

Until now, besides using Chauvenet's criterion to remove outliers, e.g.~\cite{perez_coarse_grained_2013,perez_low_energy_2016}, a human had to resolve the inconsistencies by reading the publications and trying to figure out which (if any) data set is better and remove the other one.
Unfortunately, the publications are not always accessible or they do not allow a clear statement about which data set is right or wrong.
In such a situation, we face the dilemma that rejecting one data set would be arbitrary, but feeding contradicting data sets to conventional fitting methods gives bad results.
For instance, the inclusion of two contradicting data sets leads to a larger reduction of uncertainty than if including only one.
Common sense suggests that contradictions should increase uncertainties.
\newline

The fitting method proposed in this paper resolves these issues.
Contradicting experimental data sets can be included at once.
The method automatically assigns additional uncertainties to the data sets to achieve consistency.
These additional uncertainties enable the segmentation of the experiments into consistent subsets.
A human expert can then decide upon which subset is most appropriate and feed it to a conventional fitting method.
The possibility to determine several interpretations of the data in form of consistent subsets is an advantage over Chauvenet's criterion, which yields only one interpretation.
Furthermore, overall estimates, uncertainties and correlations (in short covariance matrices) including all subsets can be obtained by means of Monte Carlo sampling.
As desired, contradictory experiment data sets increase uncertainties.
The method is mathematically well founded within the framework of Bayesian statistics.

Technicalities aside, the proposed method is related to the procedures presented in~\cite{varet_statistical_2015,varet_method_2014,varet_kriging_2013}.
These papers deal with the problem of estimating an experiment covariance matrix which may then be used to fit a model.
In contrast to that, the method introduced in this paper treats the estimation (or correction) of experiment covariance matrices as an integral part of the model fitting procedure.

The improved uncertainty quantification of evaluated nuclear data may be seen as the key feature of the proposed method.
The propagation of more realistic uncertainties of nuclear data should lead to a better assessment of simulation results and as a final consequence to safer and more efficient nuclear facilities.

\section{Method}
\subsection{Prototypic Model}
To make the discussion of the proposed method more practical, assume that we want to fit some total cross section $\sigma(E)$, which is a function of the incident energy $E$.
We take as prototypic model the function
\begin{equation}
\sigmafit(E) = \frac{
  \sum_{i=1}^M y_i \mathcal{N}\left(E\,|\,x_i,\lambda^2\right)
}{
  \sum_{j=1}^M \mathcal{N}\left(E\,|\,x_j,\lambda^2\right)
} \,.
\label{eq:protomodel}
\end{equation}
Expressions of this form appear in Nadaraya-Watson kernel regression, which is a non-parametric method for fitting.
The function $\mathcal{N}(E\,|\,x_i,\lambda^2)$ gives the probability density at location~$E$ of a normal distribution centered at energy~$x_i$ with standard deviation~$\lambda$.
The number of grid points $x_i$, their locations, and the standard deviation $\lambda$ are fixed. 
The $y_i$ are the adjustable 'model' parameters defining the shape of the function.
For notational convenience, we define the model parameter vector $\modpar = (y_1,\cdots,y_M)^T$.
Given enough grid points, the function in \cref{eq:protomodel} can mimic a multitude of possible shapes, which are determined by the choice of $\modpar$.
This prototypic model is representative for all models with a linear relationship between model parameters and predictions.
Therefore, the subsequent discussion equally applies to e.g. Fourier expansions, Legendre polynomials, and splines.
Non-linear models can be replaced by linear approximation or by a surrogate model based on a multivariate normal distribution to make them accessible for the method, e.g.~\cite{herman_development_2008}.
Of course, real physical models which possess more structure can also be used instead of series expansions.

\subsection{Standard GLS Method and Preliminaries}
The proposed method is formulated within the framework of Bayesian statistics, e.g.~\cite{jaynes_probability_2003}.
We start with outlining the popular GLS method, e.g.~\cite{capote_nuclear_2010,schnabel_modified_2017}, for nuclear data evaluation and afterward introduce modifications leading eventually to the new method.
The Bayesian update formula reads
\begin{equation}
  \rho(\modpar\,|\,\sigmaexp, \covexp) =
  \frac{
    \rho(\sigmaexp\,|\,\modpar, \covexp) \times
    \rho(\modpar \,|\,\priormodpar,\priorcovpar)
  }{
    \rho(\sigmaexp)
  } \,.
  \label{eq:stdposterior}
\end{equation}
The \textit{probability density function} (pdf) $\rho(\modpar\,|\,\priormodpar,\priorcovpar)$ reflects the prior knowledge about the model parameters.
The standard assumption is that the \textit{prior} pdf for $\modpar$ is given by a multivariate normal distribution with some center vector $\priormodpar$ and covariance matrix $\priorcovpar$, i.e. $\rho(\modpar\,|\,\priormodpar,\priorcovpar) = \mathcal{N}(\modpar\,|\,\priormodpar,\priorcovpar)$.
The \textit{likelihood} $\rho(\sigmaexp\,|\,\modpar)$ gives the probability for observing the experimental data set $\sigmaexp$ under the condition that $\vec{y}$ is the true parameter vector.
It is also given by a multivariate normal distribution $\mathcal{N}(\sigmaexp\,|\,\jacob\modpar,\covexp)$.
The covariance matrix $\covexp$ is assumed to be known a priori and reflects the statistical and systematic errors of the experiments.

The sensitivity matrix $\mat{S}$ maps the model parameters to the observables of the experiments.
It equals the Jacobian matrix, which contains the derivatives of the model predictions with respect to the model parameters.
For instance, the Jacobian matrix of the prototypic model introduced in \cref{eq:protomodel} is
\begin{equation}
	\jacobBlock{kl} = \frac{\partial}{\partial y_l} \sigmafit(E_k) =
	\frac{
  \mathcal{N}(E_k\,|\,x_l,\lambda)
	}{
  \sum_{j=1}^M \mathcal{N}(E_k\,|\,x_j,\lambda)
	} \,,
\end{equation}
where $E_k$ denotes the energy associated with the $k^\textrm{th}$ measurement point in $\sigmaexp$.
This matrix is constant with respect to the model parameters $\modpar$, which holds true for linear models in general.

The \textit{marginal likelihood} $\rho(\sigmaexp)$ yields the probability density for $\sigmaexp$ under all modelling assumptions and is determined by
\begin{equation}
	\rho(\sigmaexp) = \int
	\rho(\sigmaexp\,|\,\modpar, \covexp) \times
    \rho(\modpar \,|\,\priormodpar,\priorcovpar) \diff \modpar \,.
    \label{eq:marlikeint}
\end{equation}
It rescales the product of likelihood and prior to become a correctly normalized posterior pdf.
Due to the form of a multivariate normal pdf, conveniently expressed in terms of its logarithm,
\begin{multline}
  \ln \mathcal{N}(\vec{x}\,|\,\vec{x}_0,\mat{\Sigma}) = 
  -\frac{N}{2}\ln(2\pi) -\frac{1}{2}\ln \det{\Sigma} \\
  -\frac{1}{2} (\vec{x}-\vec{x}_0)^T\mat{\Sigma}^{-1}(\vec{x}-\vec{x}_0) \,
  \label{eq:mvnpdf}
\end{multline}
with the center vector $\vec{x}_0$ containing $N$ elements and the $N \times N$ covariance matrix $\mat{\Sigma}$, 
the marginal likelihood can be calculated analytically.
The result is (e.g.~\cite[p.~93]{bishop_pattern_2006})
\begin{equation}
  \rho(\sigmaexp) =
  \mathcal{N}(\sigmaexp\,|\,\jacob\priormodpar,\marcov) \;\;\textrm{with}\; \marcov = \jacob\priorcovpar\jacob^T + \covexp \,.
  \label{eq:stdmarlike} 
\end{equation}

Consequently, also the posterior pdf in \cref{eq:stdposterior} has a closed-form solution.
It is given by the multivariate normal distribution,
\begin{align}
  \rho(\modpar\,|\,\sigmaexp,\covexp) =& \mathcal{N}(\modpar\,|\,\postmodpar,\postcovpar)
  \;\;\textrm{with} 
  \label{eq:stdpostnorm} \\
  \postmodpar =& \priormodpar + \priorcovpar\jacob^T
  (\jacob\priorcovpar\jacob^T + \covexp)^{-1} (\sigmaexp - \jacob\priormodpar) \,, 
  \label{eq:stdpostmean} \\
  \postcovpar =& \priorcovpar - 
  \priorcovpar\jacob^T (\jacob\priorcovpar\jacob^T  + \covexp)^{-1} \jacob\priorcovpar
  \label{eq:stdpostcov} \,
\end{align}
with the new center vector $\postmodpar$ and new covariance matrix $\postcovpar$ for the model parameters.
The application of these two formulas is commonly understood as the GLS method.
Depending on the dimensions of the matrices, the following equivalent formulas may be preferred:
\begin{align}
	\postmodpar &= \postcovpar \left( \priorcovpar^{-1}\priormodpar + \jacob^T\covexp^{-1}\sigmaexp\right) \,,
  \label{eq:stdpostmeanAlt} \\
	\postcovpar &= \left(\priorcovpar^{-1}+\jacob^T\covexp^{-1}\jacob\right)^{-1}
	\label{eq:stdpostcovAlt} \,.
\end{align}
A derivation of \cref{eq:stdpostmean,eq:stdpostcov,eq:stdpostmeanAlt,eq:stdpostcovAlt} can be found in e.g.~\cite{schnabel_modified_2017}.
The method of $\chi^2$-fitting can be regarded as a special case where $\priorcovpar=\eta\mat{Q}$ with a matrix $Q$ of full rank and $\eta\to\infty$.

Defining a sensitivity matrix $\jacob_\textrm{ev}$ to map to a suitable output grid,
the result of $\jacob_\textrm{ev}\postmodpar$ together with the associated covariance matrix $\jacob_\textrm{ev}^T \postcovpar \jacob_\textrm{ev}$ enters evaluated nuclear data files.

Finally, it has to be noted that $\sigmaexp$ usually bundles data sets from different experiments $\sigmaexpPart{i}$.
Each measurement vector $\sigmaexpPart{i}$ is associated with a covariance matrix $\mat{B}_i$.
Uncertainties of distinct experiments will be assumed to be uncorrelated, which gives rise to a block diagonal structure of $\covexp$, with the $\mat{B}_i$ as blocks.
The block diagonal structure of $\mat{B}$ provides computational benefits and enables the application of the proposed method to large data sets.

\subsection{New Method}

\subsubsection{Uncertainty about the Experiment Covariance Matrix}
\label{subsubsec:expunc}

The standard GLS method assumes that the experiment covariance matrix $\mat{B}$ is perfectly known a priori.
However, in reality it is often very difficult to account exactly for all systematic uncertainties.
This circumstance suggests to regard $\mat{B}$ \textit{itself} as uncertain.
In practice, we introduce this additional uncertainty by parameterizing the covariance matrix.
More precisely, each block $\mat{B}_i$ is parametrized individually.
Among the many possibilities, an additional normalization uncertainty may be one of the most plausible options,
\begin{equation}
  \adjcovexpBlock{i}(\kappavecEl{i}) = 
  \covexpBlock{i} + \kappavecEl{i}^2 \sigmaexpPart{i}\sigmaexpPart{i}^T \,.
  \label{eq:adjcovblock}
\end{equation}
In an usual evaluation, a human evaluator would try to assign a reasonable value for the parameter $\kappavecEl{i}$ based on his knowledge about the experiment.
In the proposed method, most probable  values will be determined automatically.
Because the uncertainty about $\kappavecEl{i}$ will be expressed in terms of a probability distribution, there is lots of flexibility to account for prior knowledge.
For instance, variations of $\kappavecEl{i}$ could be restricted to take place only in a certain interval.
Noteworthy, the parametrization in \cref{eq:adjcovblock} has to be seen as a suggestion and other choices are equally reasonable.
For example, if one data set covers a broad range of incident energies, it may suit to introduce an uncertainty component that exhibits only mid-range energy correlation instead of a perfect correlation between the errors at all energies.
Such a parameterization will be demonstrated and discussed at the end of \cref{sec:demoanddiscuss}.

\subsubsection{Extended Bayesian Update Formula}
The introduction of new variables into the inference procedure necessitates an extension of the Bayesian update formula.
For convenience, we combine the variables $\kappavecEl{i}$ associated with different experiments to the vector~$\kappavec$.
The Bayesian update formula now reads
\begin{equation}
  \rho(\modpar,\kappavec \,|\, \sigmaexp) =
  \frac{
  \rho(\sigmaexp \,|\, \modpar, \kappavec) \times
  \rho(\modpar \,|\,\priormodpar,\priorcovpar) \times 
  \rho(\kappavec) 
  }{
  \rho(\sigmaexp)
  } \,.
  \label{eq:extbayesformula}
\end{equation}
As in the case of the standard GLS, the likelihood is given by a multivariate normal distribution, $\rho(\sigmaexp \,|\, \modpar, \kappavec) = \mathcal{N}(\sigmaexp\,|\,\jacob\modpar,\adjcovexp(\kappavec))$.
Noteworthy, the covariance matrix $\covexp$ is replaced by $\adjcovexp(\kappavec)$ whose blocks are determined by \cref{eq:adjcovblock}.
The specification of the prior for the model parameters $\rho(\modpar \,|\,\priormodpar,\priorcovpar) = \mathcal{N}(\modpar\,|\,\priormodpar,\priorcovpar)$  mirrors the standard GLS method.
We postpone discussing the choice of the prior pdf $\rho(\kappavec)$ for a moment.

Contrary to the standard GLS approach, the marginal likelihood
\begin{multline}
  \rho(\sigmaexp) = \int \left(
  \int
  \rho( \modpar, \sigmaexp \,|\, \kappavec) \times
  \rho(\kappavec) 
   \diff \modpar
  \right)  
  \diff \kappavec \\
  \textrm{with}\;
  \rho( \modpar, \sigmaexp \,|\, \kappavec) = 
  \rho(\sigmaexp \,|\, \modpar, \kappavec) \times
  \rho(\modpar \,|\,\priormodpar,\priorcovpar)
  \label{eq:newmarpost}
\end{multline}
has no straight-forward analytical solution.
Only the inner integral can be analytically evaluated.
Noting that it has the same form as \cref{eq:marlikeint}, the solution analogous to \cref{eq:stdmarlike} is 
\begin{multline}
   \rho(\kappavec \,|\, \sigmaexp) \propto \rho(\sigmaexp, \kappavec) =
  \mathcal{N}(\sigmaexp \,|\, \jacob\priormodpar, \;
  \marcov) \times \rho(\kappavec) \\ 
  \;\;\textrm{with}\;\; 
  \marcov = \jacob\priorcovpar\jacob^T + \adjcovexp(\kappavec) \,.
  \label{eq:kappamarpost}
\end{multline}
Because this expression is proportional to the posterior pdf $\rho(\kappavec \,|\, \sigmaexp)$, it is the key to assess the consistency of the experiment data sets.
The vector $\kappavec'$ that maximizes \cref{eq:kappamarpost} contains the most probable values for the parameters in the experiment covariance matrix.
It tells us which data sets are consistent and which are not, and how wrong the inconsistent ones are estimated to be.
In the case of several local maxima, each maximum is associated with a certain interpretation of the experiments.
Details concerning the computation and optimization of $\rho(\kappavec \,|\, \sigmaexp)$ will be discussed in \cref{subsubsec:effcomp}

\subsubsection{Choice of the Shape of the Prior Distribution}
\label{subsubsec:choicepriorpdf}
For a full specification of $\rho(\kappavec \,|\, \sigmaexp)$, we have to define the prior pdf $\rho(\kappavec)$.
Knowledge about correlations between different $\kappavecEl{i}$ is usually limited.
Furthermore, the automated detection of most probable adjustments of the experiment covariance matrix is one of the main reasons for the introduction of the new method.
Consequently, we want to avoid an informative prior for the covariance matrix parameters. 

The information content of a pdf can be characterized in terms of \textit{entropy} (e.g.~\cite{cover_elements_1991})---the higher the entropy of a pdf, the lower the information content.
Given only the marginal pdfs $\rho(\kappavecEl{i}), i=1..N$ with associated entropies $\mathcal{H}\big[\rho(\kappavecEl{i})\big]$, the joint pdf $\rho(\kappavec)$ with highest entropy and compatible with all the marginal pdfs is just the product of the marginal pdfs,
\begin{equation}
	\rho(\kappavec) = \rho(\kappavecEl{1}) \rho(\kappavecEl{2}) \dots \rho(\kappavecEl{N}) \,.
\end{equation}
This result follows from the subadditivity of the entropy \cite[p.~28]{cover_elements_1991},
\begin{multline}
	\mathcal{H}\big[ \rho(\kappavecEl{1}), \rho(\kappavecEl{2}), \cdots, \rho(\kappavecEl{N}) \big]
	\leq \\
	\mathcal{H}\big[\rho(\kappavecEl{1})\big] +
	\mathcal{H}\big[\rho(\kappavecEl{2})\big] +
	... +
	\mathcal{H}\big[\rho(\kappavecEl{N})\big] \,,
\end{multline}
with equality only if the variables $\kappavecEl{i}$ are statistically independent, i.e. the joint pdf factorizes into the product of the marginal pdfs.

Concerning the functional form of $\rho(\kappavecEl{i})$, I investigated the Laplace pdf, the normal pdf, and the improper uniform pdf in a schematic evaluation of the neutron-proton total cross section.
The term improper refers to the fact that the uniform pdf extends over the complete real line and hence cannot be normalized.
Details about the findings will be presented in \cref{sec:demoanddiscuss}.
However, some results must be already anticipated here in order to provide a complete picture of the method.

In my investigation, I found arguments in favor of the Laplace pdf,
\begin{equation}
  \mathcal{L}(\kappavecEl{i}\,|\,\delta_i) = 
  \frac{1}{\sqrt{2}\,\delta_i}
  \exp\left( -\frac{\sqrt{2}\,|\kappavecEl{i}|}{\delta_i} \right) \,.
  \label{eq:Laplacepdf}
\end{equation}
This pdf is symmetric with mean zero and standard deviation~$\delta_i$.
Experiments believed to be more correct could be associated with smaller $\delta_i$ than those being more distrusted.
However, in an automated evaluation without much human involvement, there is no reason to favor one experiment over another a priori.
Therefore, I investigated only the case where all $\delta_i$ are equal.

In the studied scenario, the experiment data in combination with the uniform distribution did not sufficiently constrain the posterior pdf.
Even though the most probable assignments $\kappavec'$ were usually reasonable, the relative standard deviations of the $\kappavecEl{i}$ exceeded thousand percent---unreasonably large.
In contrast to that, both the normal pdf and the Laplace pdf with a reasonable standard deviation $\delta$ restricted sufficiently the spread of $\rho(\kappavec \,|\, \sigmaexp)$.
Noteworthy, the Laplace pdf tended to set more $\kappavecEl{i}$ to zero at the cost of slightly increased values of non-zero parameters.
I regard this behavior to favor sparse solutions as beneficial.
If this behavior is not desired, the normal distribution should be prefered.

\begin{figure}[b]
\centering
\includegraphics{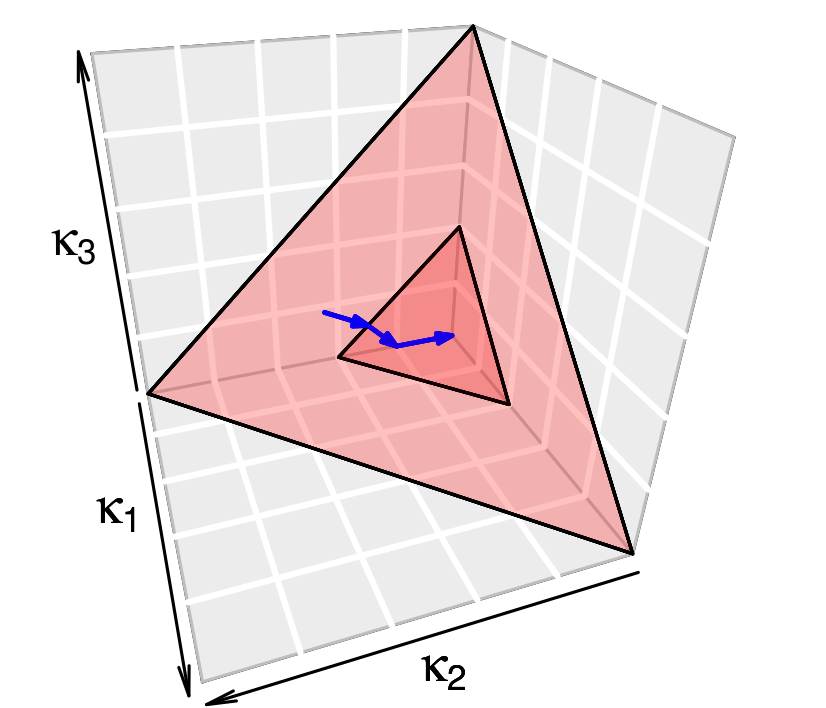}
\caption{
Surfaces of equal probability for a product of three identical Laplace pdfs.
Displayed is the octant where all parameters have positive values.
The blue arrows show an exemplary path of steepest ascent. 
}
\label{fig:LaplacePriorVis}
\end{figure}
In fact, the logarithm of the product of identical Laplace pdfs appears (up to a constant) as penalty term in LASSO regression~\cite{tibshirani_regression_1996} where it serves the exact purpose of variable elimination.
To understand this behaviour, consider the set $\big\{ \kappavec \,|\, \tau = \rho(\kappavec) \big\}$ for a fixed positive real number $\tau$ and with $\rho(\kappavec)$ being specified as a product of identical Laplace pdfs.
The vectors in this set define a hypercube whose corners are aligned with the parameter axes.
The gradient perpendicular to the surface of this hypercube does almost everywhere \textit{not} point exactly to the origin of the coordinate system, as it would be the case for the product of identical normal distributions.
Instead, following locally the direction of steepest ascent leads to a vector with one component being zero, say $\kappavecEl{1}=0$.
Continuing on the path of steepest ascent leads stepwise to the elimination of more and more parameters.
The center of the distribution is reached only at the very end of the path.
The process is visualized in \cref{fig:LaplacePriorVis}.
This theoretical argument explains why the preference for sparse solutions is a general feature when using a product of Laplace pdfs as prior pdf.

\subsubsection{Choice of the Parameter $\delta$}
\label{subsubsec:deltachoice}
The prior pdf~$\rho(\kappavec)$ of the covariance matrix parameters~$\kappavec$ depends itself on parameters.
In the last section we encountered the standard deviation $\delta$ as the parameter defining the shape of the identical Laplace pdfs.
We can make this dependence explicit by writing $\rho(\kappavec\,|\,\delta$) instead of $\rho(\kappavec)$.
Having introduced a new parameter, which value should we assign to it?

Again, we anticipate some results from \cref{sec:demoanddiscuss}
If the $\kappavecEl{i}$ denote normalization uncertainties, then $\delta$ should be set to a plausible value for the normalization uncertainty.
So if we think it is quite probable that some experiments have normalization errors between $5\%$ and $10\%$ which were not considered in the original experiment covariance matrix $\covexp$, then $\delta$ should be also in that range.
It appears that evaluation results are only mildly dependent on the exact choice of $\delta$.
This approach, however, is rather subjective.
Next we discuss data-driven approaches to alleviate the problem of subjectivity.

A sufficient criterion to determine whether the value of~$\delta$ is large enough is to calculate the generalized $\chi^2$-value
\begin{equation}
	\chi^2(\kappavec') = 
	(\sigmaexp-\jacob\priormodpar)^T
	\left[\marcov(\kappavec')\right]^{-1}
	(\sigmaexp-\jacob\priormodpar)
	\label{eq:genchisq}
\end{equation}
where $\kappavec'$ maximizes $\rho(\kappavec \,|\, \sigmaexp)$, see \cref{eq:kappamarpost}.
The quantity $\chi^2(\kappavec')/N$, with $N$ being the number of measurement points, should be close to one.
Otherwise one or more of the following statements is true: 1) the value of $\delta$ is too low, 2) the model to fit the data is misspecified, 3) a normalization uncertainty is not enough to correct the misspecified experiment covariance matrix.
At the end of \cref{sec:demoanddiscuss} we discuss besides a normalization uncertainty also a more flexible energy-dependent uncertainty.

An approach that directly aims at the determination of $\delta$ is the maximization of the  marginal likelihood $\rho(\sigmaexp)$ defined in \cref{eq:newmarpost}. 
Because the prior $\rho(\kappavec\,|\,\delta)$ is conditioned on $\delta$, we more appropriately write $\rho(\sigmaexp\,|\,\delta)$ instead of $\rho(\sigmaexp)$.
The resulting value represents the probability density to obtain the measurement vector $\sigmaexp$ given a certain value of~$\delta$.
Of course, this probability density is also conditioned on the assumption of the model, the parameterization of the adjusted experiment covariance matrix $\adjcovexp$, and the prior specifications of all occurring parameters
Selecting a value for $\delta$ that maximizes $\rho(\sigmaexp \,|\, \delta)$ is a sensible choice, effectively removing subjectivity.

Unfortunately, it seems as there is no analytical expression for $\rho(\sigmaexp \,|\, \delta)$.
We can approximately solve the integral by using Monte Carlo integration in combination with \textit{importance sampling}, e.g. \cite[p.~131]{rubinstein_simulation_1981}.
The idea is to identically rewrite \cref{eq:newmarpost} as 
\begin{equation}
	\rho( \sigmaexp \,|\, \delta) = 
	\int \frac{
		\rho(\sigmaexp\,|\,\kappavec) \times \rho(\kappavec\,|\,\delta)
	}{
		\phi(\kappavec)
	} 
	\times \phi(\kappavec) \diff \kappavec \,.
\end{equation}
Given that $\phi(\kappavec)$ is a pdf from which we can draw a sample~$\kappavec_1,\kappavec_2,\cdots, \kappavec_K$,
an estimate of this integral is
\begin{equation}
	\rho( \sigmaexp \,|\, \delta) \approx
	\frac{1}{K} \sum_{i=1}^{K} 
	\frac{
		\rho(\sigmaexp\,|\,\kappavec_i) 
	}{
		\phi(\kappavec_i)
	} \times \rho(\kappavec_i\,|\,\delta) \,.
	\label{eq:newmarapprox}
\end{equation}
The choice of $\phi(\kappavec)$ will be discussed in a moment.

Finding the value of $\delta$ that maximizes $\rho( \sigmaexp \,|\, \delta)$ means to evaluate the integral for many possible values of $\delta$.
In order to scan the parameter space in a systematic way, we can evaluate the integral on a grid of reasonable values $\delta_1, \delta_2, \cdots, \delta_M$.
In fact, we can use the same sequence~$\kappavec_1,\kappavec_2,\cdots, \kappavec_K$ drawn from $\phi(\kappavec)$ to estimate all the integrals $\phi(\sigmaexp \,|\, \delta_i), i=1..M$ in parallel.
The ratios $\rho(\sigmaexp\,|\,\kappavec_i)/\phi(\kappavec_i)$ in \cref{eq:newmarapprox} are the same for all integrals.
Only the last factor $\rho(\kappavec_i\,|\,\delta)$ has to be recomputed for each value $\delta_i$.
Because it does not depend on the experiment data sets and is of a simple form, e.g. a Laplace pdf, it can be evaluated quickly.

How should the sampling distribution $\phi(\kappavec)$ be chosen?
A sampling distribution that declines at a faster rate in the tails than the other part of the integrand destroys convergence in importance sampling.
In order to protect against this case, we define a mixture of possible posterior pdfs associated with the values $\delta_j$ on the grid,
\begin{equation}
	\phi(\kappavec) = \mathcal{I} \times \mathcal{N}(\sigmaexp \,|\, \jacob\priormodpar, \; \marcov(\kappavec))
	\times 
	\sum_j \rho(\kappavec \,|\, \delta_j) \,.
	\label{eq:mixlaplace}
\end{equation}
We emphasize the dependence of $\marcov$ on $\kappavec$, see \cref{eq:kappamarpost}.
The normalization constant $\mathcal{I}$ is not required to generate samples if using the \textit{Metropolis-Hastings} (MH) algorithm~\cite{hastings_monte_1970} (see also the appendix).

The fact that the pdf $\phi(\kappavec)$ enters \cref{eq:newmarapprox} and hence estimates of $\rho(\sigmaexp\,|\,\delta_m)$ depend on $\mathcal{I}$ is not important.
Since the normalization $\mathcal{I}$ affects each estimate in the same way, its value does not influence the relative likelihoods.

Due to the form of \cref{eq:mixlaplace} and due to $\rho(\sigmaexp\,|\,\kappavec_i)=\mathcal{N}(\sigmaexp \,|\, \jacob\priormodpar, \; \marcov(\kappavec))$, the estimate in \cref{eq:newmarapprox} simplifies to
\begin{equation}
	\rho( \sigmaexp \,|\, \delta_m) \approx
	\frac{1}{\mathcal{I} K} \sum_{i=1}^{K} \omega_m(\kappa_i)
	\label{eq:marlikeestdepdelta}
\end{equation}
with the abbreviation
\begin{equation}
	\omega_m(\kappa_i) =
	\frac{
		\rho(\kappavec_i\,|\,\delta_m)
	}{
		\sum_j \rho(\kappavec_i \,|\, \delta_j)
	} 
	\,.
	\label{eq:miximpweight}
\end{equation}

The value $\delta_{m'}$ associated with the biggest value $\rho(\sigmaexp\,|\,\delta_{m'})$ should be selected for the analysis.

\subsubsection{Overall Prediction and Covariance Matrix}
Besides finding consistent subsets of experiments, which is a question of maximizing $\rho(\sigmaexp\,|\,\delta)$ given in \cref{eq:kappamarpost}, we may be interested in an overall prediction and the associated covariance matrix by averaging over all possible interpretations.
Technically, to find the overall prediction, we have to solve
\begin{equation}
  \begin{split}
  \hat{\modpar} = \mathds{E}[ \modpar ] &= \int \int \modpar \, \rho(\modpar,\kappavec \,|\, \sigmaexp) 
  \diff \modpar \diff \kappavec \\
  &= \int \left(  \int \modpar \, \rho(\modpar \,|\, \kappavec, \sigmaexp) \diff \modpar \right) \,
  \rho(\kappavec \,|\, \sigmaexp) 
  \diff \kappavec \,.
  \end{split}
  \label{eq:overallmean1}
\end{equation}
The inner integral yields the expectation of $\modpar$ under $\rho(\modpar \,|\, \kappavec, \sigmaexp)$.
This conditioned posterior pdf has the same functional form as the posterior pdf of the standard GLS method in \cref{eq:stdpostnorm}.
For the latter distribution we know the result of the integral, which is \cref{eq:stdpostmean}.
Therefore, the result of the inner integral in \cref{eq:overallmean} is given by
\begin{equation}
  \postmodpar(\kappavec) = \priormodpar + \priorcovpar\jacob^T
  \left(\jacob\priorcovpar\jacob^T + \adjcovexp(\kappavec)\right)^{-1} 
  (\sigmaexp - \jacob\priormodpar) \,.
  \label{eq:GLSmeanC}
\end{equation}
If the computation of the inverse matrix is infeasible, the form of \cref{eq:stdpostmeanAlt} can be used.
Using this analytic expression, the integral in \cref{eq:overallmean1} takes the form
\begin{equation}
	\hat{\modpar} = \mathds{E}[ \modpar ] =
	\int \postmodpar(\kappavec) \, \rho(\kappavec  \,|\, \sigmaexp) \diff \kappavec \,.
	\label{eq:overallmean}
\end{equation}
Likely no analytic solution exists for this remaining integral and we have to take recourse to Monte Carlo integration.

The simplification of the integral for the overall covariance matrix follows analogous steps.
The overall covariance matrix can be written as
\begin{equation}
\begin{split}
  \hat{\mat{\Sigma}} &=
  \mathds{E}[ \modpar \modpar^T ] - \mathds{E}[ \modpar ] \mathds{E}[ \modpar^T ] \\  
  &= \int   
  \left(\modpar\modpar^T - \hat{\modpar}\hat{\modpar}^T \right)
  \rho(\modpar,\kappavec \,|\, \sigmaexp)
  \diff \modpar \diff \kappavec \\
  &= \int \left(  \int \modpar\modpar^T 
   \, \rho(\modpar \,|\, \kappavec, \sigmaexp) \diff \modpar  \right) \,
  \rho(\kappavec \,|\, \sigmaexp) \diff \kappavec 
	- \hat{\modpar}\hat{\modpar}^T  
  \,.
\end{split}
\label{eq:overallcov1}
\end{equation}
Using the identity
\begin{equation}
	\postcovpar = \int \modpar\modpar^T \,
	\rho(\modpar \,|\, \kappavec, \sigmaexp)
	\diff \modpar \,
	- \postmodpar(\kappavec)\postmodpar(\kappavec)^T
	\label{eq:stdpostcovint}
\end{equation}
whose solution is analogous to the standard GLS method, see \cref{eq:stdpostcov},
\begin{equation}
	\postcovpar(\kappavec) = \priorcovpar - 
	\priorcovpar\jacob^T \left(\jacob\priorcovpar\jacob^T  + \adjcovexp(\kappavec)\right)^{-1} \jacob\priorcovpar \,,
\end{equation}
we can express \cref{eq:overallcov1} as
\begin{equation}
 \hat{\mat{\Sigma}} =
 \int \left(\postcovpar(\kappavec) + \postmodpar(\kappavec)\postmodpar(\kappavec)^T \right)
  \, \rho(\kappavec  \,|\, \sigmaexp) \diff \kappavec - \hat{\modpar}\hat{\modpar}^T \,.
  \label{eq:overallcov}
\end{equation}
As for the overall prediction, also this integral likely has no analytic solution.

Consequently, the integrals for the overall prediction in \cref{eq:overallmean} and the overall covariance matrix in \cref{eq:overallcov} have to be solved by means of Monte Carlo integration.
One possibility is to obtain a sample $\kappavec_1,\cdots,\kappavec_K$ from $\rho(\kappavec \,|\, \sigmaexp)$ using the Metropolis-Hastings algorithm (see the appendix) and to approximate the integrals in terms of mean values.
The approximations for the overall prediction and covariance matrix are then
\begin{align}
	\hat{\modpar} &\approx 
	\frac{1}{K}	\sum_{i=1}^K \postmodpar(\kappavec_i) \,,
	\;\textrm{and}\;\; \\
	\hat{\mat{\Sigma}} &\approx 
	\frac{1}{K} \sum_{i=1}^K 
	\left(\postcovpar(\kappavec) + \postmodpar(\kappavec)
	\postmodpar(\kappavec)^T \right) 
		- \hat{\modpar}\hat{\modpar}^T
	 \,.
\end{align}

However, if we have determined the most likely standard deviation $\delta$ for the multivariate Laplace prior according to the sampling procedure outlined in \cref{subsubsec:deltachoice}, there is an alternative route.
We can reuse the samples $\kappa_1,\cdots,\kappa_K$ drawn from the mixture pdf $\phi(\kappavec)$ specified in \cref{eq:mixlaplace}.
Using the notation $\omega_m(\kappa_i)$ introduced in \cref{eq:miximpweight}, the approximations are given by
\begin{align}
	\hat{\modpar} &\approx 
	\frac{1}{\mathcal{J}K} \sum_{i=1}^K \omega_m(\kappa_i) \, \modpar(\kappavec_i) \,,
	\;\textrm{and} \\
	\hat{\mat{\Sigma}} &\approx
	\frac{1}{\mathcal{J}K} 
	\sum_{i=1}^K \omega_m(\kappa_i) \,
	\left(\postcovpar(\kappavec) + \postmodpar(\kappavec)\postmodpar(\kappavec)^T \right) 
	- \hat{\modpar}\hat{\modpar}^T \,.
\end{align}
The index $m$ refers to the value $\delta_m$ that has been selected as the most likely candidate.
The unknown normalization constant $\mathcal{J}$ can be estimated by
\begin{equation}
 \mathcal{J} = \frac{1}{K} \sum_{i=1}^{K} \omega_m(\kappa_i) \,.
\end{equation}
Please note that this normalization constant is not identical to $\mathcal{I}$ of \cref{eq:miximpweight} because it is also determined by the unknown normalization of the posterior pdf $\rho(\kappavec \,|\, \sigmaexp)$. 

The described scheme of approximation is known as \textit{self-normalized importance sampling} in the statistics literature, e.g.~\cite[p.~131]{rubinstein_simulation_1981}.
The approach to solve some integrations of a multi-dimensional integral analytically and to use Monte Carlo sampling to evaluate the remaining integrals is termed as \textit{Conditional Monte Carlo} in~\cite[p.~125]{rubinstein_simulation_1981}.

\subsubsection{Efficient Computation}
\label{subsubsec:effcomp}
The identification of plausible covariance matrix parameters $\kappavec$ (e.g. normalization uncertainties) is a question of maximizing $\rho(\kappavec \,|\, \sigmaexp) \propto \rho(\sigmaexp\,|\,\kappavec)\times \rho(\kappavec)$ given in \cref{eq:kappamarpost}.
Also determining an overall prediction and the associated covariance matrix involves the evaluation of this pdf.
The prior pdf $\rho(\kappavec)$ can be calculated quickly if opting for a product of Laplace or normal pdfs.
Contrary to that, the computation of the likelihood $\rho(\sigmaexp\,|\,\kappavec) = \mathcal{N}(\sigmaexp \,|\, \jacob\priormodpar, \;
  \marcov)$ may be computationally expensive.
Inspecting the form of this multivariate normal pdf,
\begin{multline}
  \ln \mathcal{N}\big(\sigmaexp \,|\,
   \jacob\priormodpar, \marcov(\kappavec)\big) = 
  -\frac{N}{2}\ln(2\pi) -\frac{1}{2}\ln \det{\marcov(\kappavec)} \\
  -\frac{1}{2} (\sigmaexp-\jacob\priormodpar)^T\left(\mat{\marcov(\kappavec)}\right)^{-1}
  (\sigmaexp-\jacob\priormodpar)
  \label{eq:marlike2}
\end{multline}
with $\marcov(\kappavec) = \jacob\priorcovpar\jacob^T + \adjcovexp(\kappavec)$,
we see that the expensive operations are the calculation of the determinant and the inversion of the matrix $M$. The dimension of this matrix is determined by the total number of experiment data points.
The time to invert a $10^4\times 10^4$ matrix may be tens of seconds on a contemporary personal computer.
In addition, numerical maximization and the generation of a Monte Carlo chain require at least thousands of function evaluations.
Clearly, to \textit{efficiently} compute $\rho(\sigmaexp\,|\,\kappavec)$ is important.
This section explains therefore the efficient computation of $\ln \mathcal{N}(\sigmaexp \,|\, \jacob\priormodpar,\;\marcov)$ and its gradient $\diff(\ln \rho(\kappavec \,|\, \sigmaexp)) / \diff \kappavec$.
Having an analytic expression for the gradient offers great benefits in numerical maximization.

Inverse matrices will often appear in the discussion, so we use the notation $\tilde{\mat{X}}$ instead of $\mat{X}^{-1}$ to save space.
Further, we just write $\adjcovexp$ and $\marcov$ from now on, but the dependence on $\kappavec$ should be kept in mind.
Using the Woodbury identity (\cref{apx:eq:woodbury} in the appendix), we express the inverse of $\marcov$ as
\begin{equation}
  \invmarcov = 
  \invadjcovexp - \invadjcovexp\jacob\left(
  \invpriorcovpar + \jacob^T\invadjcovexp\jacob
  \right)^{-1}\jacob^T\invadjcovexp \,.
  \label{eq:effinvmarcov}
\end{equation}

The measurement vector $\sigmaexp$ is partitioned into sub-vectors $\sigmaexpPart{i}$ associated with different experiments.
For each $\sigmaexpPart{i}$ there is a sensitivity matrix $\jacobBlock{i}$ to map from model parameters to the respective predictions.
Therefore, the sensitivity matrix is partitioned into $\jacob = (\jacobBlock{1}^T,\cdots,\jacobBlock{N}^T)^T$.
Exploiting this partitioned form and the the block diagonal structure of $\invadjcovexp$ allows us to write
\begin{equation}
  \jacob^T\invadjcovexp\jacob = 
  \sum_k \jacobBlock{k}^T\invadjcovexpBlock{k}\jacobBlock{k} \,.
  \label{eq:STinvCS}
\end{equation}
Because the number of data points in one data set is usually limited, say less than hundred, the computation of the inverse matrices $\invadjcovexpBlock{k}$ can be performed fast on contemporary personal computers.
The sum of matrices in the brackets in \cref{eq:effinvmarcov} leads to a matrix of the same dimension as $\invpriorcovpar$, hence it is determined by the number of model parameters.
I expect models or series expansions not to have more than hundreds of adjustable parameters.

The inverse matrix $\invmarcov$ appears only in the matrix product $\vec{u}^T\invmarcov\vec{u}$ with $\vec{u} = \sigmaexp - \jacob\priormodpar$.
Also $\vec{u}$ is partitioned into sub-vectors $\vec{u}_i = \sigmaexpPart{i} - \jacobBlock{i}\priormodpar$.
Noting that
\begin{equation}
  \invadjcovexp\vec{u} =
  \left(
  \left(\invadjcovexpBlock{1}\vec{u}_1\right)^T,\cdots,
  \left(\invadjcovexpBlock{N}\vec{u}_N\right)^T
  \right)^T
  \label{eq:partCu}
\end{equation}
is a vector and considering the form of \cref{eq:effinvmarcov}, we see that $\vec{u}^T\invmarcov\vec{u}$ can be completely evaluated in terms of computationally cheap matrix-vector products.

To tackle the determinant, we use the matrix determinant lemma~(\cref{apx:eq:matrixdeterminantlemma} in the appendix) to obtain
\begin{equation}
\begin{split}
  \ln |\marcov| &=
  \ln |\invpriorcovpar + \jacob^T\invadjcovexp\jacob| +
  \ln |\adjcovexp| + \ln|\priorcovpar| = \\
  &=
  \ln \left\vert\invpriorcovpar + 
  \sum_k \jacobBlock{k}^T\invadjcovexpBlock{k}\jacobBlock{k}
  \right\vert +
  \sum_k \ln |\invadjcovexpBlock{k}| + \ln |\invpriorcovpar| \,,
\end{split}
\end{equation}
with $|\mat{X}|$ being the short-hand notation for $\det{\mat{X}}$.
To get from the first to the second line, we used \cref{eq:STinvCS} and the fact that the determinant of a block diagonal matrix is the product of the block determinants.
As elaborated above, determinants have to be taken only from comparatively low dimensional matrices.

Finally, we briefly discuss how to calculate the gradient of $\ln \rho(\kappavec \,|\, \sigmaexp)$.
Blocks of $\invmarcov$ are given by
\begin{equation}
  \invmarcovBlock{ij} = 
  \delta_{ij}\invadjcovexpBlock{i} 
  - \invadjcovexpBlock{i}\jacobBlock{i}\left(
  \invpriorcovpar + \sum_k \jacobBlock{k}^T\invadjcovexpBlock{k}\jacobBlock{k}
  \right)^{-1}\jacobBlock{j}^T\invadjcovexpBlock{j} \,,
  \label{eq:blocksofM}
\end{equation}
where $\delta_{ij} = 1$ for $i=j$ and $\delta_{ij}=0$ for $i\neq j$.
The derivative of $\ln |\mat{M}|$ in \cref{eq:marlike2} can be written as (\cref{apx:eq:logdetderivative} in the appendix)
\begin{equation}
  \frac{\partial \ln |\mat{M}|}{\partial\kappavecEl{i}} = 
  \tr\left[
    \invmarcov
    \frac{\partial \adjcovexp}{\partial \kappavecEl{i}}
  \right]
  = 
    \tr\left[ \invmarcovBlock{ii}
    \frac{\partial \adjcovexpBlock{i}}{\partial \kappavecEl{i}}
    \right] \,.
\end{equation}
The partial derivatives of the matrix product with respect to the parameters $\kappavecEl{i}$ are (\cref{apx:eq:invmatderiv} in the appendix)
\begin{equation}
  \frac{\partial \vec{u}^T\invmarcov\vec{u}}{\partial \kappavecEl{i}} =
  -\vec{u}^T \invmarcov
  \frac{\partial\adjcovexp}{\partial\kappavecEl{i}}
  \invmarcov \vec{u} =
  -\vec{u}^T \invmarcovBlock{.i}
  \frac{\partial\adjcovexpBlock{i}}{\partial\kappavecEl{i}}
  \invmarcovBlock{i.} \vec{u} \,.
\end{equation}
A point in the index of a matrix denotes the inclusion of all rows or columns.
Again, exploiting the partitioned form in \cref{eq:partCu}, the evaluation of this quantity only involves computationally inexpensive matrix-vector products.
The inner derivative completes the determination of the gradient.
For the normalization uncertainty defined in \cref{eq:adjcovblock}, we get
\begin{equation}
  \frac{\partial\adjcovexpBlock{i}}{\partial\kappavecEl{i}} =
  2 \kappavecEl{i} \left( \sigmaexpPart{i}\sigmaexpPart{i}^T \right) \,.
\end{equation}

Now, equipped with an analytic expression for the gradient $(\diff/\diff \kappavec) (\ln \mathcal{N}\big(\sigmaexp \,|\,
   \jacob\priormodpar, \marcov(\kappavec)\big))$, the full gradient $\diff(\ln \rho(\kappavec \,|\, \sigmaexp))/\diff\kappavec$ is straight-forward to compute.
Considering the complete log-posterior pdf
\begin{equation}
  \ln \rho(\kappavec \,|\, \sigmaexp) \stackrel{C}{=} \ln \rho(\sigmaexp\,|\,\kappavec)
  + \ln \rho(\kappavec) \,,
\end{equation}
we just have to add $\diff(\ln \rho(\kappavec))/\diff \kappavec$.
In the case of identical Laplace pdfs, see \cref{eq:Laplacepdf}, the components of the latter gradient are
\begin{equation}
	\frac{\partial \ln\rho(\kappavec)}{\partial \kappavecEl{i}} =
	-\frac{\sqrt{2}}{\delta} \textrm{sign}(\kappavecEl{i})
\end{equation}
with $\textrm{sign}(\kappavecEl{i})$ being either $-1$ or $+1$ according to the sign of $\kappavecEl{i}$.

In summary, this section elaborated on the efficient computation of $\ln \rho(\kappavec \,|\, \sigmaexp)$ and its gradient.
The inversion and the determinant of the potentially large matrix $\marcov$ have been transformed to the same operations on the comparatively small matrix
$\left(\invpriorcovpar + \jacob^T\invadjcovexp\jacob\right)$.
The size of the latter matrix is determined by the number of model parameters.
Due to the block-diagonal structure of $\adjcovexp$, the inversion can be performed fast and yields another block-diagonal matrix.
The resulting matrix  $\invadjcovexp$ only enters an inexpensive matrix-vector product whose evaluation profits again from the block-diagonal structure of~$\invadjcovexp$.
For the same reasons, the analytic expression of the gradient can be also computed quickly.

Finally, the availability of the gradient enables the application of gradient-based optimization algorithms, such as the \textit{BFGS} ~\cite{broyden_convergence_1970} or \textit{L-BFGS} algorithm~\cite{byrd_limited_1995}.
Especially the latter algorithm is very memory efficient and hence suited for a scenario with many experiment data sets.
As another useful feature, it allows the specification of parameter boundaries.

\FloatBarrier
\section{Demonstration and Discussion}
\label{sec:demoanddiscuss}
\begin{figure}[t]
	\centering
	\includegraphics{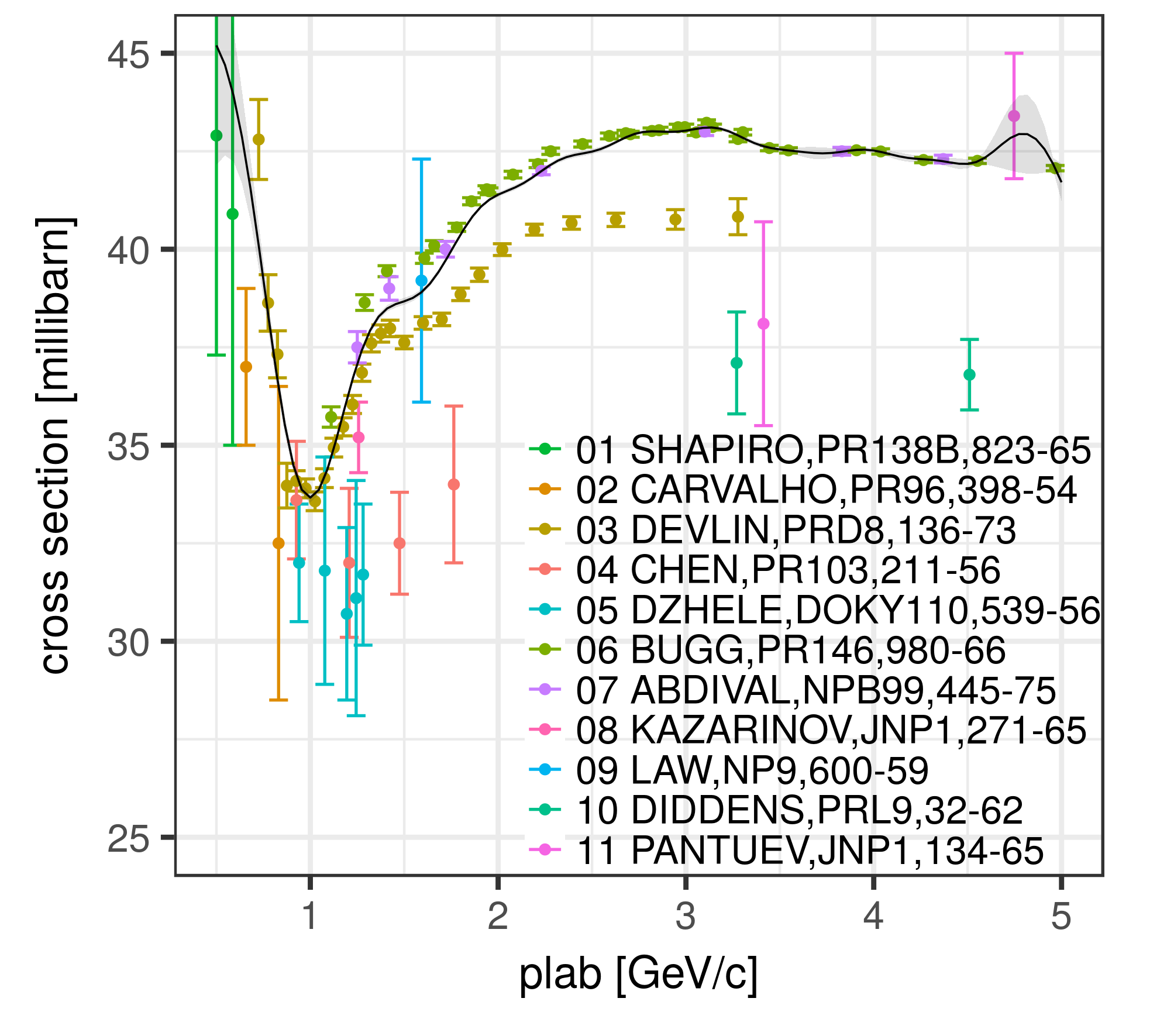}
	\caption{Experiment data used in the schematic evaluation.
	The black line is the resulting prediction from a GLS fit of the original data without any correction of the uncertainty assumptions.}
	\label{fig:expdata}
\end{figure}

The method will be demonstrated at the example of eleven data sets taken from \cite{landolt_total_1988} with measurements of the proton-neutron total cross section.
I selected the data points with incident momenta (in the laboratory frame) between 0.5\,GeV/c and 5\,GeV/c because many discrepant data sets are available in this range.
I assume that only statistical uncertainties are present, which leads to a diagonal matrix $\covexp$.
Correlated errors, such as an uncertainty about the detector efficiency, could also be included, but the respective information is not always available in nuclear databases.
The experiment data are shown in \cref{fig:expdata}.
Considering the extent of the error bars indicating the 68\% confidence interval, the data are clearly inconsistent.

The series expansion introduced in \cref{eq:protomodel} with fifty expansion terms provides the model to fit the data.
The exact specification employed in this section is given by
\begin{equation}
\sigmafit(E) = \frac{
  \sum_{i=1}^{50} y_i \mathcal{N}\left(E\,|\,x_i,\lambda^2\right)
}{
  \sum_{i=1}^{50} \mathcal{N}\left(E\,|\,x_i,\lambda^2\right)
} 
\end{equation}
with $\lambda=0.2$ and $x_j=0.2+i\times (5-0.5)/50$.
This model imposes a certain degree of smoothness on the cross section curve but besides that can adapt flexibly to the data.

The prior $\rho(\modpar \,|\,\priormodpar,\priorcovpar)$ for the parameter vector $\modpar$ is a multivariate normal distribution $\mathcal{N}(\modpar \,|\,\priormodpar,\priorcovpar)$ with all elements in $\priormodpar$ equal forty.
The associated prior covariance matrix $\priorcovpar$ is diagonal with all elements equal thousand.
This prior covers well the experiment data.

Using the standard GLS method to fit the model yields the curve illustrated in \cref{fig:expdata}.
The 68\% error band is hardly visible at most energies.
Further, the fit runs in between the data sets around 1.5\,GeV/c and the associated uncertainty band excludes them.
This observation is associated with the presence of inconsistent data.
The result of the GLS method serves as a reference to which the results of the proposed method can be compared.

In the first part of the demonstration, blocks of the adjusted covariance matrix $\adjcovexp(\kappavec)$ are parameterized as
\begin{equation}
  \adjcovexpBlock{i}(\kappavecEl{i}) = 
  \covexpBlock{i} + \kappavecEl{i}^2 \sigmaexpPart{i}\sigmaexpPart{i}^T \,.
  \label{eq:demo_adjcovexp}
\end{equation}
This parameterization introduces an additional normalization uncertainty $\kappavecEl{i}$ for each experiment data set.
Therefore, the vector $\kappavec$ contains eleven variables.
Afterward, the method will be also applied with a more flexible energy-dependent parameterization.

Because normalization uncertainties $\kappavecEl{i}$ are themselves uncertain, we need to specify a prior pdf $\rho(\kappavec\,|\,\delta)$.
I tested the method with the following three prior specifications:
\begin{align}
	\rho_\textrm{U}(\kappavec) &= \textrm{const} \,,
	\label{eq:demo_unifprior} \\
	\rho_\textrm{N}(\kappavec\,|\,\delta) &=
	\prod_{i=1}^{11} \frac{1}{\sqrt{2\pi}\,\delta} 
	\exp\left( 
	-\frac{1}{2} \frac{\kappavecEl{i}^2}{\delta^2}\right)	 \,,
	\label{eq:demo_normalprior} \\
	\rho_\textrm{L}(\kappavec\,|\,\delta) &= 
	\prod_{i=1}^{11} \frac{1}{\sqrt{2}\,\delta}
  \exp\left( -\frac{\sqrt{2}\,|\kappavecEl{i}|}{\delta} \right) \,.
  	\label{eq:demo_laplaceprior}
\end{align}
The first pdf is an improper uniform pdf.
Using this prior pdf, the posterior pdf $\rho(\kappavec\,|\,\sigmaexp)$ is exclusively determined by the marginal likelihood $\rho(\sigmaexp\,|\,\kappavec)$.
The second pdf is a product of identical normal distribution and the third pdf a product of identical Laplace pdfs.
The parameter $\delta$ signifies in both cases the standard deviation of the distribution.

\subsection{Selection of $\delta$}
\label{subsec:selectdelta}
In order to carry out the method with either $\rho_\textrm{N}(\kappavec\,|\,\delta)$ or $\rho_\textrm{L}(\kappavec\,|\,\delta)$, a suitable $\delta$ has to be selected.
Linked to these prior pdfs are the following mixture pdfs: 
\begin{align}
	\phi_\textrm{N}(\kappavec) = \mathcal{I}_\textrm{N} \times 
	\mathcal{N}(\sigmaexp \,|\, \jacob\priormodpar, \; \marcov(\kappavec))
	\times 
	\sum_{j=1}^{30} \rho_\textrm{N}(\kappavec\,|\,\delta_j) \,, \\
	\phi_\textrm{L}(\kappavec)
	 = \mathcal{I}_\textrm{L}
	 \times \mathcal{N}(\sigmaexp \,|\, \jacob\priormodpar, \; \marcov(\kappavec))
	\times 
	\sum_{j=1}^{30} \rho_\textrm{L}(\kappavec\,|\,\delta_j) \,.
	\label{eq:lapmixpdf}
\end{align}
The functional form of these pdfs was introduced in \cref{eq:mixlaplace}.
The components are characterized by $\delta_j = 0.01\times j$ and the normalization constants $\mathcal{I}_\textrm{N}, \mathcal{I}_\textrm{L}$ are set to one.
Considering \cref{fig:expdata}, the appropriate value of $\delta$ is somewhere between 1\% and 30\% and hence the form of the mixture pdfs justified.

Next, a sample of each mixture has to be obtained by means of the Metropolis-Hastings algorithm.
I employed $\psi(\kappavec'\,|\,\kappavec) = \mathcal{N}(\kappavec'\,|\,\kappavec, \tau^2 \mathds{1})$ as proposal pdf.
After tentative runs of the MH~algorithm with different values of $\tau$, the assignment $\tau=0.045$ turned out to be a good choice yielding acceptance rates around 30\% for both mixture pdfs.
Unless otherwise stated, this proposal distribution is employed throughout the demonstration.
In principle, the choice of $\tau$ could be automated, too, e.g. \cite{haario_adaptive_2001}.
Investigation in this respect is left as future work.

\begin{figure}[t]
	\centering
	\includegraphics{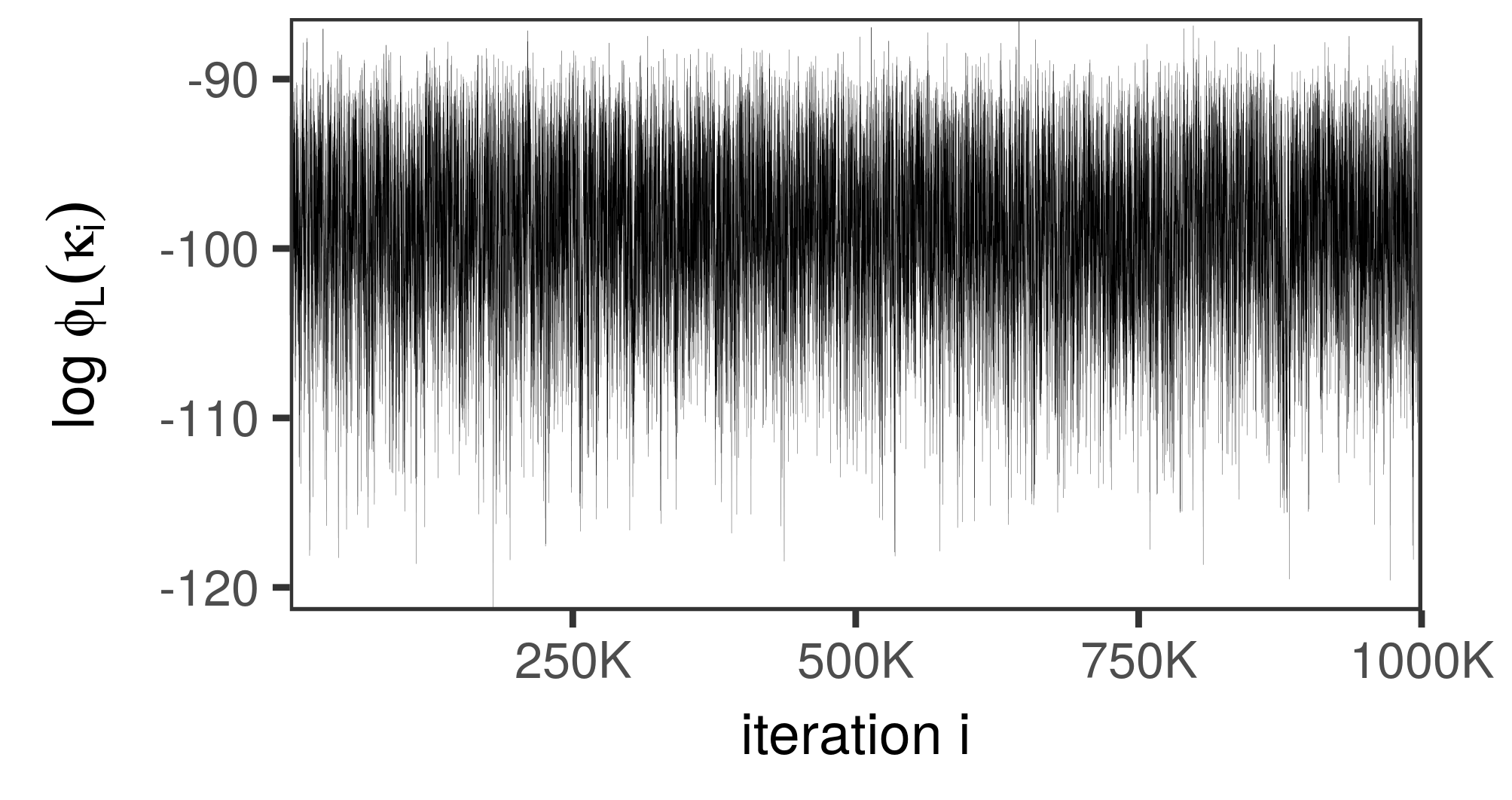}
	\caption{Evolution of $\log \phi_\textrm{L}(\kappavec)$, see \cref{eq:lapmixpdf}, in the process of MH sampling.}
	\label{fig:mixLapLogDens}
\end{figure}

Finally, I created for each mixture pdf a Monte Carlo chain with one million vectors.
The evolution of $\log\phi_\textrm{L}(\kappavec)$ as a function of the iteration count  is illustrated in \cref{fig:mixLapLogDens}.
No obvious drift can be noticed, which gives evidence that the MH chain represents a sample from $\phi_\textrm{L}(\kappavec)$.
The density evolution plot for $\log\phi_\textrm{N}(\kappavec)$ looked similar without any sign of drifting. 

Calculating the marginal likelihood $\rho( \sigmaexp \,|\, \delta_j)$ according to \cref{eq:marlikeestdepdelta} for all $\delta_j$ in the mixture pdf $\phi_\textrm{L}(\kappavec)$, we learn that the maximum appears at $\delta_\textrm{L}=0.13$.
Hence, this value should be used in the procedure.
The respective value in the case of $\phi_\textrm{N}(\kappavec)$ was $\delta_\textrm{N}=0.11$.

\begin{figure}[t]
	\centering
	\includegraphics{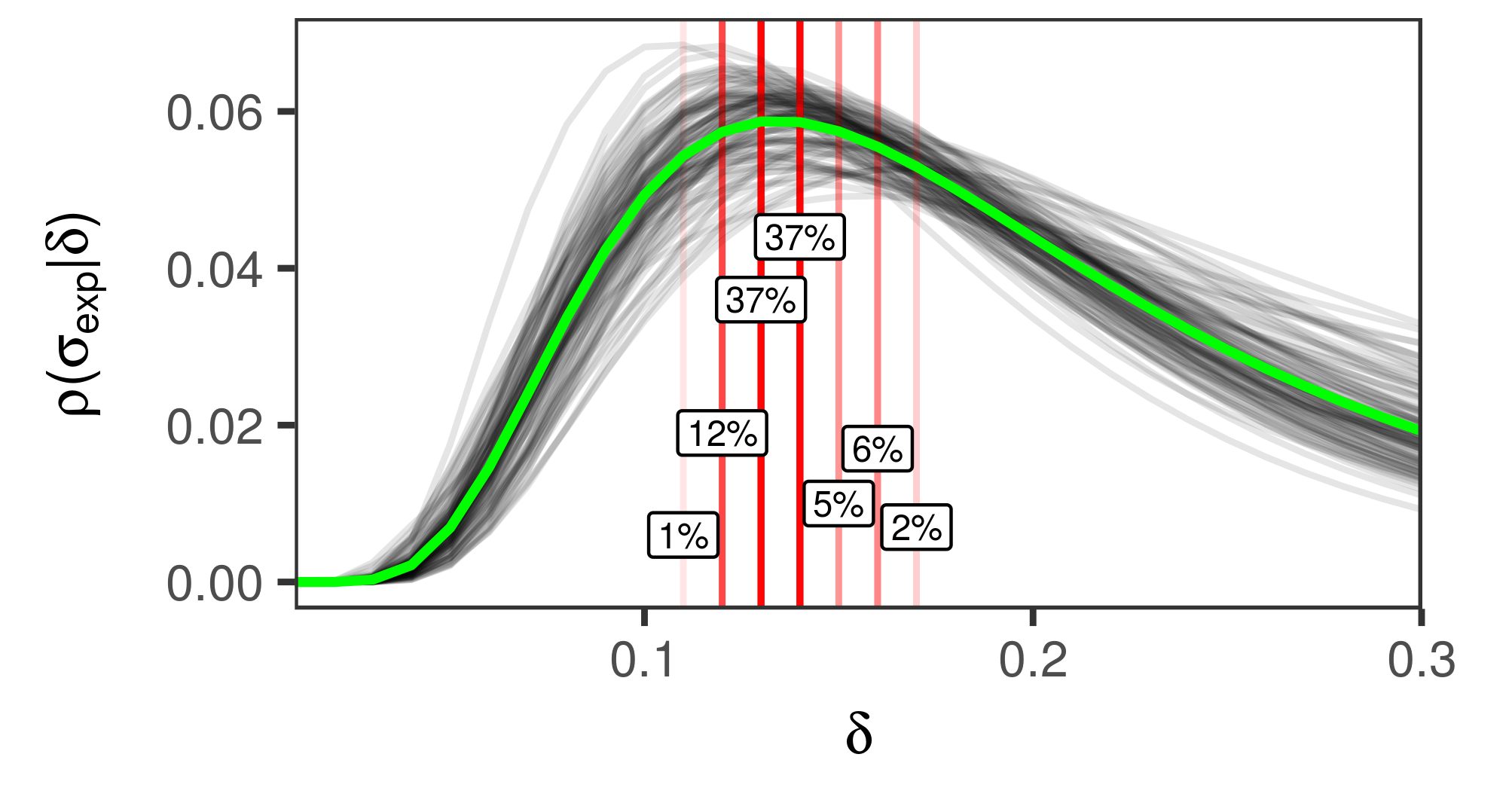}
	\caption{Fluctuations in the estimate of the parameter $\delta$ that maximizes $\rho_\textrm{L}(\sigmaexp\,|\,\delta)$.
	Red vertical lines and overlaid percentage numbers indicate the proportion of cases that led to the estimate at the respective value of $\delta$.	
	  }
	\label{fig:mixLapDispMax}
\end{figure}

One may be concerned that the peak is rather flat and the relative likelihoods associated with $\delta$ values in vicinity are similar, as visualized by the green line in \cref{fig:mixLapDispMax}.
This observation begs the question of how strong the position of the maximum fluctuates if estimated from a smaller chain.
To address this question, I cut the MC chain into chunks consisting of $10^4$ vectors and estimated the relative likelihoods and the position of the maximum on the basis of each chunk.
The ensemble of black curves in \cref{fig:mixLapDispMax} conveys an impression of the variations in the relative likelihoods.
The red vertical lines denote which $\delta$ values were identified as maxima according to the chunks.
The overlaid percentages display the proportion of chunks with the respective location as maximum.
In spite of the rather large fluctuations of the relative likelihoods, the maximum of $\delta$ was estimated to be either $0.13$ or $0.14$ in $74\%$ of the cases.
In all cases, the maximum was situated between $0.11$ and $0.17$.
At first glance, this rather large spread suggests to always construct long chains---a time-consuming process.
For example, the generation of one million vectors took about six hours on a contemporary personal computer.
However, one reason for the large spread is the insensitivity of the likelihood $\rho(\sigmaexp\,|\,\kappavec)$ to the choice of $\delta$.
This feature implies that the mean vector $\hat{\modpar}$ calculated from the posterior pdf according to \cref{eq:overallmean} is also rather insensitive to the exact value of $\delta$.
Thus, the fluctuations of a few percent are acceptable and a MC chain with $10^4$ vectors seems (at least in the studied example) sufficient.
Further evidence for the validity of this statement will be provided in the next sections.

\subsection{Segmenting Experiment Data into Subsets}
\label{subsec:demo_segdata}
Consistent subsets of data can be found by maximizing the marginal posterior pdf $\rho(\kappavec\,|\,\sigmaexp)$.
The three prior pdfs introduced in \cref{eq:demo_unifprior,eq:demo_normalprior,eq:demo_laplaceprior} lead to the following posterior pdfs:
\begin{align}
	\rho_\textrm{U}(\kappavec\,|\,\sigmaexp) &= \mathcal{I}_\textrm{U} \times 
	\mathcal{N}(\sigmaexp \,|\, \jacob\priormodpar, \; \marcov(\kappavec))
	\times 
	\rho_\textrm{U}(\kappavec) \,, \\
	\label{eq:demo_unifpost}
	\rho_\textrm{N}(\kappavec\,|\,\sigmaexp) &= \mathcal{I}_\textrm{N} \times 
	\mathcal{N}(\sigmaexp \,|\, \jacob\priormodpar, \; \marcov(\kappavec))
	\times 
	\rho_\textrm{N}(\kappavec\,|\,\delta_\textrm{N}) \,, \\
	\rho_\textrm{L}(\kappavec\,|\,\sigmaexp)
	 &= \mathcal{I}_\textrm{L}
	 \times \mathcal{N}(\sigmaexp \,|\, \jacob\priormodpar, \; \marcov(\kappavec))
	\times 
 	\rho_\textrm{L}(\kappavec\,|\,\delta_\textrm{L}) \,.
\end{align}
The normalization constants $\mathcal{I}_{\textrm{U}},\mathcal{I}_{\textrm{N}},\mathcal{I}_{\textrm{L}}$ are not required for the maximization.
We can conveniently set them to one.
The values $\delta_\textrm{N}=0.11$ and $\delta_\textrm{L}=0.13$ are taken from the previous section.

Because the posterior probability densities usually cover a huge range, which easily leads to a numerical over- or underflow, the numerical maximization was carried out for the logarithms of these pdfs.
The L-BFGS-B algorithm~\cite{byrd_limited_1995} as implemented by the \textit{optim} function in the statistical programming language \textit{R}~\cite{r_development_core_team_r_2008} proved to be reliable.
This algorithm takes advantage of an analytic expression of the gradient, see the derivation starting from \cref{eq:blocksofM}.
Furthermore, it permits the specification of box constraints.
Box constraints can be used to effectively deal with proper uniform priors.
Yet, more important for our case, we can constrain parameters to be positive and exclude zero as solution.
Taking into account the form of \cref{eq:demo_adjcovexp} and how it enters the multivariate normal pdf in \cref{eq:marlike2}, we recognize the posterior pdfs to be symmetric around $\kappavec = \vec{0}$.
This insight justifies the restriction to positive values.
The exclusion of zero is important in the case of $\rho_\textrm{L}(\kappavec\,|\,\sigmaexp)$ as the gradient exhibits a discontinuity if some $\kappavecEl{i}=0$.
The L-BFGS-B algorithm relies on the gradient to be continuous, and thus discontinuities potentially cause problems.

Due to these reasons, I constrained all $\kappavecEl{i}$ to lie between $1\times 10^{-4}$ and $5\times 10^{-1}$.
The upper bound protects against unreasonable solutions with normalization uncertainties greater than $50\%$.
Concerning the overall setup of the L-BFGS-B algorithm, I limited the maximal number of iterations in the numerical maximization procedure to thousand and specified that the Hessian matrix should be estimated based on the precedent twenty iteration steps.
For each maximization attempt, the vector $\kappavec$ was initialized with values drawn uniformly from the range between $1\times 10^{-4}$ and $5\times 10^{-1}$.
Each maximization attempt was repeated ten times to ensure that the global maximum has been indeed found.
\begin{table*}[ht]
  \centering
\newcommand{\tbf}[1]{\textbf{#1}}
\begin{tabular}{llllllllllllllr}\toprule
	& $\delta$ & $\ell$ &
	$\kappavecEl{1}$ & $\kappavecEl{2}$ & $\kappavecEl{3}$ & 
	$\kappavecEl{4}$ & $\kappavecEl{5}$ & $\kappavecEl{6}$ & 
	$\kappavecEl{7}$ & $\kappavecEl{8}$ & $\kappavecEl{9}$ & 
	$\kappavecEl{10}$ & $\kappavecEl{11}$  &
	$\chi^2 / N$
\\ \midrule

	GLS & --- & --- &
	.00 & .00 & .00 & .00 & .00 & .00 & .00 &
 	.00 & .00 & .00 & .00 & 16.13 \\ 
	$\rho_\textrm{U}$ & --- & --- &
	\tbf{.00} & .04 & .10 & \tbf{.00} & \tbf{.00} & .14 &
	.13 & .06 & .10 & \tbf{.00} & .13 & 0.73 \\
	$\rho_\textrm{N}$ & .11 & --- &
	\tbf{.00} & .03 & .07 & \tbf{.00} & \tbf{.00} & .10 &
	.09 & .04 & .05 & \tbf{.00} & .09 & 0.79 \\
	$\rho_\textrm{L}$ & .13 & $1.0\times 10^0$  & 
	\tbf{.00} & \tbf{.00} & .07 & \tbf{.00} & \tbf{.00} &
	.09 & .09 & .03 & \tbf{.00} & \tbf{.00} & .08 & 0.82 \\
	$\rho_\textrm{L}$ & .13 & $1.8\times 10^{-3}$ & 
	\tbf{.00} & .09 & [\tbf{.00}] & .07 & .07 & .03 & .03 &
	.01 & \tbf{.00} & .07 & \tbf{.00} & 0.98 \\
	$\rho_\textrm{L}$ & .13 & $1.2\times 10^{-5}$ & 
	\tbf{.00} & .12 & .03 & .10 & .10 & [\tbf{.00}] & \tbf{.00} &
	.06 & \tbf{.00} & .10 & \tbf{.00} & 1.07 \\
	$\rho_\textrm{L}$ & .17 & --- &
	\tbf{.00} & \tbf{.00} & .07 & \tbf{.00} & \tbf{.00} & .10 &
	.09 & .04 & \tbf{.00} & \tbf{.00} & .09 & 0.81
\\
\bottomrule
\end{tabular}

  \caption{Posterior maxima $\kappavec$ based on the prior distributions specified in \cref{eq:demo_unifprior,eq:demo_normalprior,eq:demo_laplaceprior}.
For the pdfs $\rho_\textrm{N}$ and $\rho_\textrm{L}$, results based on different values of $\delta$ are presented.
Square brackets denote that the respective $\kappavecEl{i}$ was fixed at zero.
The index $i$ refers to the experiment data set, see \cref{fig:expdata}.
The value $\chi^2/N$ is the result of \cref{eq:genchisq} divided by the number of data points.
Relative likelihoods $\ell$ are stated for the case $\rho_\textrm{L}$ with $\delta=0.13$. 
}   
  \label{tab:demo_segdata}
\end{table*}

The results of the numerical maximization are summarized in \cref{tab:demo_segdata}.
The solutions based on $\rho_\textrm{U}$ and $\rho_\textrm{N}$ are comparable in structure.
The same $\kappavecEl{i}$ are set to zero, only the remaining $\kappavecEl{i}$ are pulled closer to zero in the case of $\rho_\textrm{N}$.
Albeit the slightly more constrained solution, the $\chi^2/N$ value associated with $\rho_\textrm{N}$ is only a bit larger and still below one.
This observation suggests that $\rho_\textrm{N}$ should be preferred over $\rho_\textrm{U}$.

Comparing $\rho_\textrm{N}$ and $\rho_\textrm{L}$, we can make the interesting observation that more parameters $\kappavecEl{i}$ are set to zero in the case of $\rho_\textrm{L}$ despite the standard deviation $\delta_\textrm{L}$ being larger than $\delta_\textrm{N}$.
\Cref{fig:LaplacePriorVis} visualized the reason for this behavior.
The non-zero parameters of the two solutions are very similar.
Further, both solutions are consistent because their $\chi^2/N$ values are close to one.
Ockham's principle states that among the many explanations compatible with a certain observation, the explanation with the least assumptions should be chosen.
According to this principle, the pdf $\rho_\textrm{L}$ should be preferred over $\rho_\textrm{N}$ because it favors sparse solutions.

\Cref{fig:mixLapDispMax} showed the fluctuations in the determination of~$\delta_\textrm{L}$.
To study the sensitivity of the maximum $\kappavec$ to the choice of $\delta_\textrm{L}$, I performed the maximization also for $\rho_\textrm{L}$ with $\delta_L=0.17$.
The result displayed in \cref{tab:demo_segdata} is hardly  different from the result based on $\rho_\textrm{L}$ with $\delta_L=0.13$.
Consequently, as already conjectured in the previous section, the fluctuations of $\delta$ do not significantly alter the results.

The discussion so far provided arguments in favor of $\rho_\textrm{L}(\kappavec)$ as prior pdf.
Please note, however, that all prior specifications led to acceptable values of $\chi^2/N$.
If the $\chi^2$ value really comes from a $\chi^2$-distribution, then its standard deviation is $\sigma = \sqrt{2N}$.
Because there are 84 experiment data points in total, we get $\sigma/N$ = 0.15.
The associated $95\%$ interval $[0.7, 1.3]$ includes all observed $\chi^2/N$ values.
This fact indicates that the method is effectively able to \textit{correct} the uncertainty assumptions of the experiments.
Contrary to that, the GLS fit of the uncorrected data is associated with the too large value $\chi^2/N=16.13$.

\begin{figure}[t]
	\centering
	\includegraphics{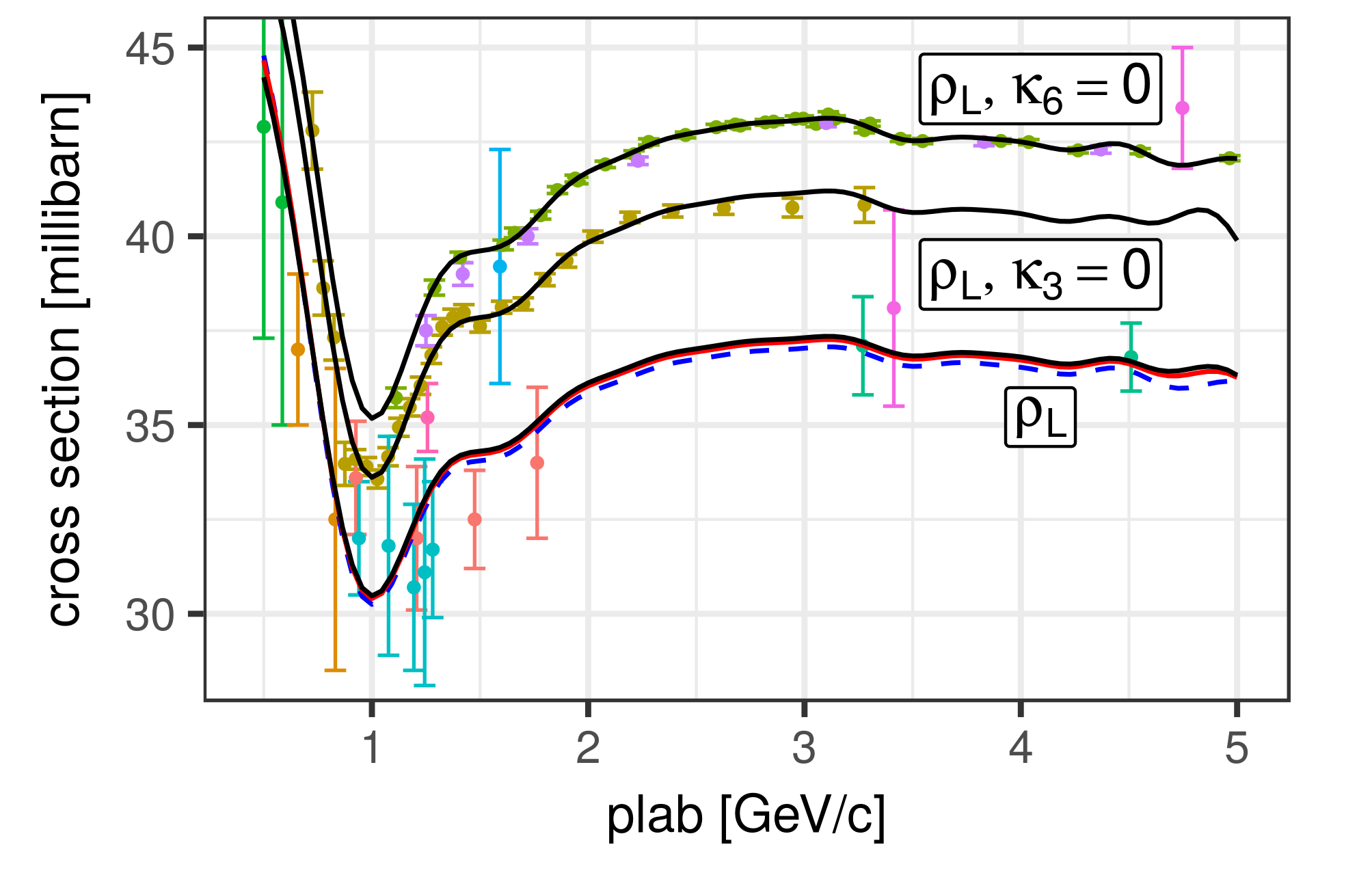}
	\caption{The black line labelled with $\rho_\textrm{L}$ shows the posterior maximum associated with the prior in \cref{eq:demo_laplaceprior} with $\delta=0.13$.
	The blue dashed line and the red solid line in vicinity are the maxima of $\rho_\textrm{U}$ and $\rho_\textrm{N}$ with $\delta=0.11$.	
	The upper two black lines are the maxima of $\rho_\textrm{L}$ with $\delta=0.13$ under the constraint that $\kappa_3=0$ and $\kappa_6=0$, respectively.
	}
	\label{fig:demoseg}
\end{figure}

The \textit{segmentation of data sets} can be performed by removing data sets whose normalization constants exceed a certain threshold.
We can assume that one (or several) data sets are correct and fix their $\kappavecEl{i}$ at zero during the optimization.
\Cref{tab:demo_segdata} shows this constrained optimization for $\rho_\textrm{L}$ with $\delta=0.13$ and either $\kappavecEl{3}=0$ or $\kappavecEl{6}=0$ fixed.
The predictions of the GLS method, see \cref{eq:GLSmeanC}, using the obtained vectors $\kappavec$ are depicted in \cref{fig:demoseg}.
In contrast to the GLS fit of the uncorrected data shown in \cref{fig:expdata}, these fits agree well with the data sets that were \textit{a priori} assumed to be correct.

Finally, one mode of failure must be mentioned.
If a normalization uncertainty for each data set is not sufficient to achieve consistency among data sets, we may also get unfavorable results with the new approach.
For example, just as in the standard GLS fit of the uncorrected data, the prediction could run in between the data sets and exhibit too small uncertainties.

Nevertheless, even in this scenario, we may still use the method in combination with normalization uncertainties as exploration technique.
Even though a normalization uncertainty would not be enough to make the data sets consistent, very likely the method would still identify inconsistent data sets by introducing large normalization uncertainties.

Another option is to use a more flexible parametrization of the covariance matrix $\adjcovexp(\kappavec)$, which is able to model more elaborate uncertainty assumptions.
This approach will be discussed and demonstrated in \cref{subsec:beyondnormalization}.

\subsection{Overall Prediction and Covariance matrix}
The additional layer of uncertainty about the normalization uncertainties~$\kappavecEl{i}$ also increases the uncertainty in the final estimates of the model parameters.
The posterior pdf $\rho(\kappavec\,|\,\sigmaexp)$ provides the probability density for any choice of~$\kappavec$.
Each of these choices produces a different result in the GLS method.
Therefore, the overall prediction is calculated as a weighted mean of all these results, see~\cref{eq:overallmean}.
Analogously, also the overall covariance matrix is---loosely speaking---the weighted mean of the covariance matrices conditioned on the different choices of $\kappavec$, see \cref{eq:overallcov}.

\begin{figure}[b]
	\centering
	\includegraphics{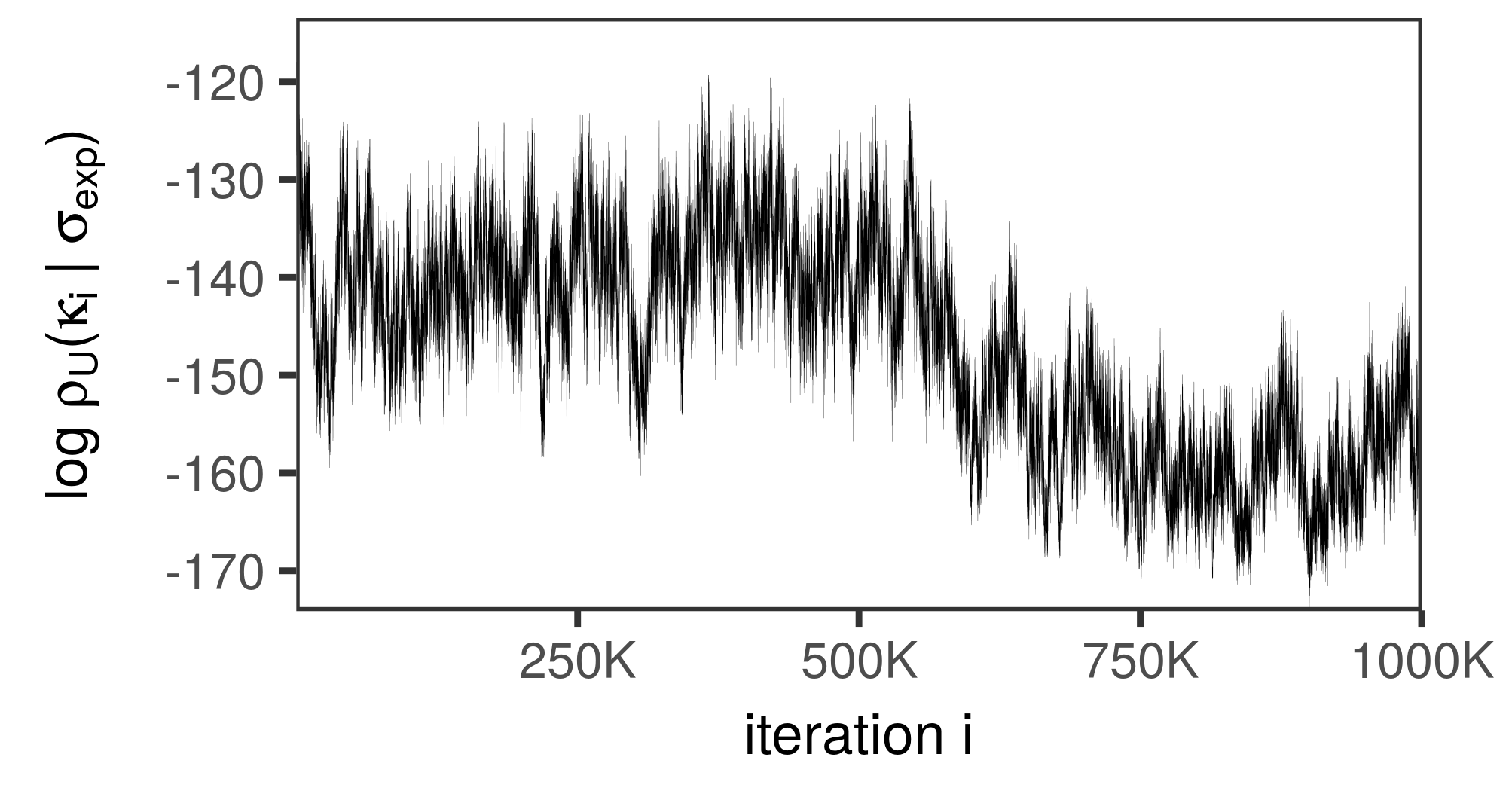}
	\caption{Evolution of $\phi_\textrm{U}(\kappavec\,|\,\sigmaexp)$, see \cref{eq:demo_unifpost}, in the process of MH sampling. }
	\label{fig:uniformLogDens}
\end{figure}

It turned out that the uniform prior $\rho_\textrm{U}$ did not lead to reasonable solutions.
The MC chain to draw samples from $\rho_\textrm{U}(\kappavec\,|\,\sigmaexp)$ given in \cref{eq:demo_unifpost} did not reach its stationary distribution even after one million iterations.
The employed proposal pdf $\psi(\kappavec'\,|\,\kappavec) = \mathcal{N}(\kappavec'\,|\,\kappavec, \tau^2 \mathds{1})$ with $\tau=0.3$ lead to acceptance rate of $30\%$ and more.
The evolution of the chain is illustrated in \cref{fig:uniformLogDens}.
I performed several runs of the MH algorithm, but the observed behavior persisted.
The parameters $\kappavecEl{i}$ acquired values in the order of hundred and sometimes even thousand.
Normalization uncertainties of 10000\% are clearly absurd in our scenario.
Taking into account that the maximum of $\rho_\textrm{U}(\kappavec\,|\,\sigmaexp)$ shown in \cref{tab:demo_segdata} was reasonable, I conclude that the data alone do not sufficiently constrain the normalization uncertainties.
Technically, the determinant in \cref{eq:marlike2} responsible for the decline of the probability density does not grow fast enough with increasing $\kappavecEl{i}$.
Inspecting plots of the probability density as a function of the parameters $\kappavecEl{i}$ confirmed this hypothesis.

\begin{figure}[t]
	\centering
	\includegraphics{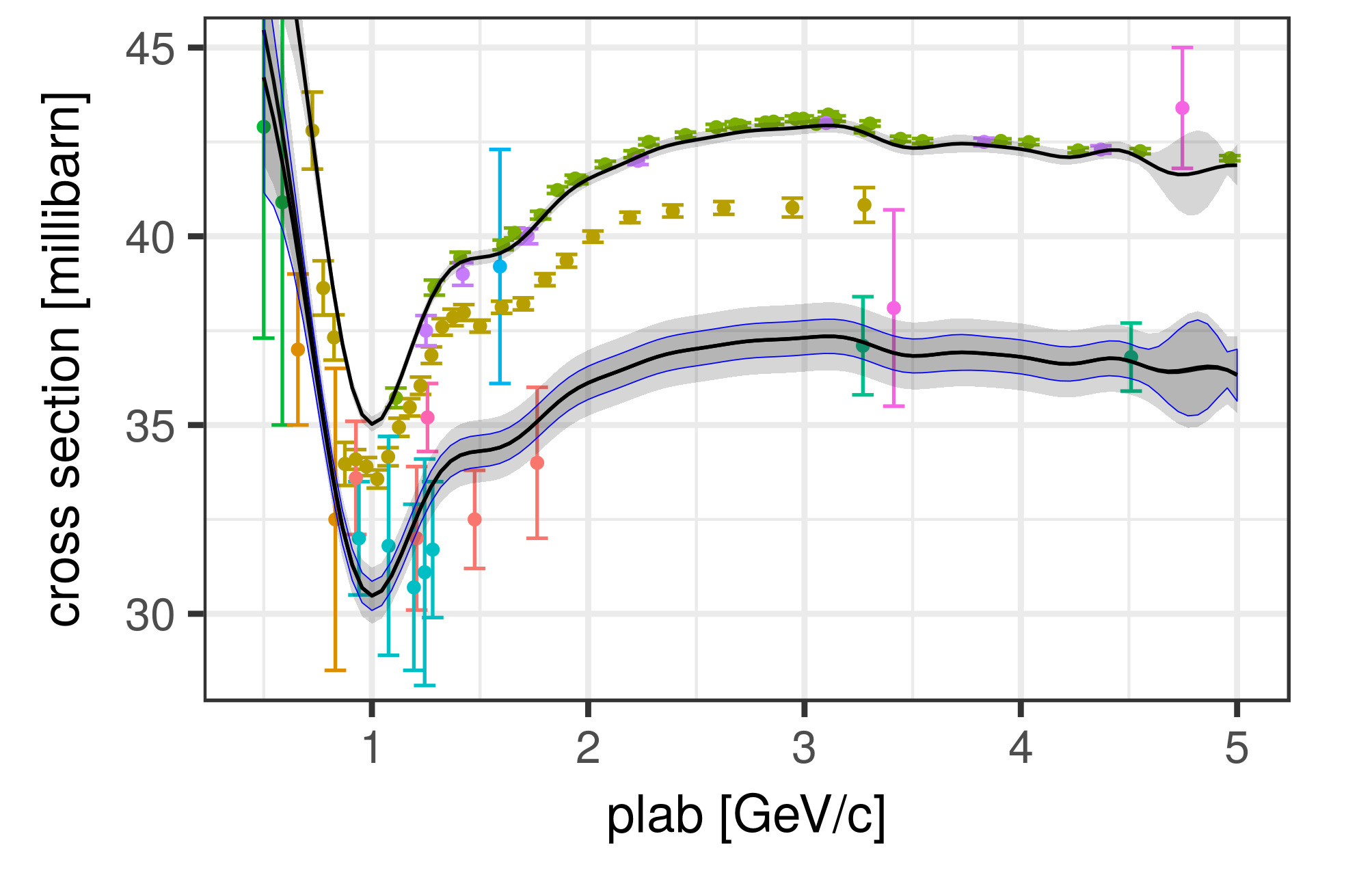}
	\caption{Lower curve shows the overall prediction and associated $1\sigma$-confidence band.
	The light-blue edge indicates the extent of uncertainty linked to the posterior maximum prediction.
	The upper curve shows the overall prediction under the constraint $\kappavecEl{6}=0.005$.} 
	\label{fig:demoovpred}
\end{figure}

The chains to draw samples from $\rho_\textrm{N}(\kappavec\,|\,\sigmaexp)$ with $\delta_\textrm{N}=0.11$ and $\rho_\textrm{L}(\kappavec\,|\,\sigmaexp)$ with $\delta_\textrm{L}=0.13$ behaved well.
Because their predictions and associated uncertainties were visually indistinguishable, I only discuss the case of $\rho_\textrm{L}$.
The overall prediction is illustrated in \cref{fig:demoovpred}.
Interestingly, it coincides with the prediction associated with the posterior maximum, compare with  \cref{fig:demoseg}.
Only the $1\sigma$ uncertainty band of the overall prediction is larger than that one of the posterior maximum prediction, which is the expected behavior due to averaging over the covariance matrices.

\begin{figure}[t!]
	\centering
	\includegraphics{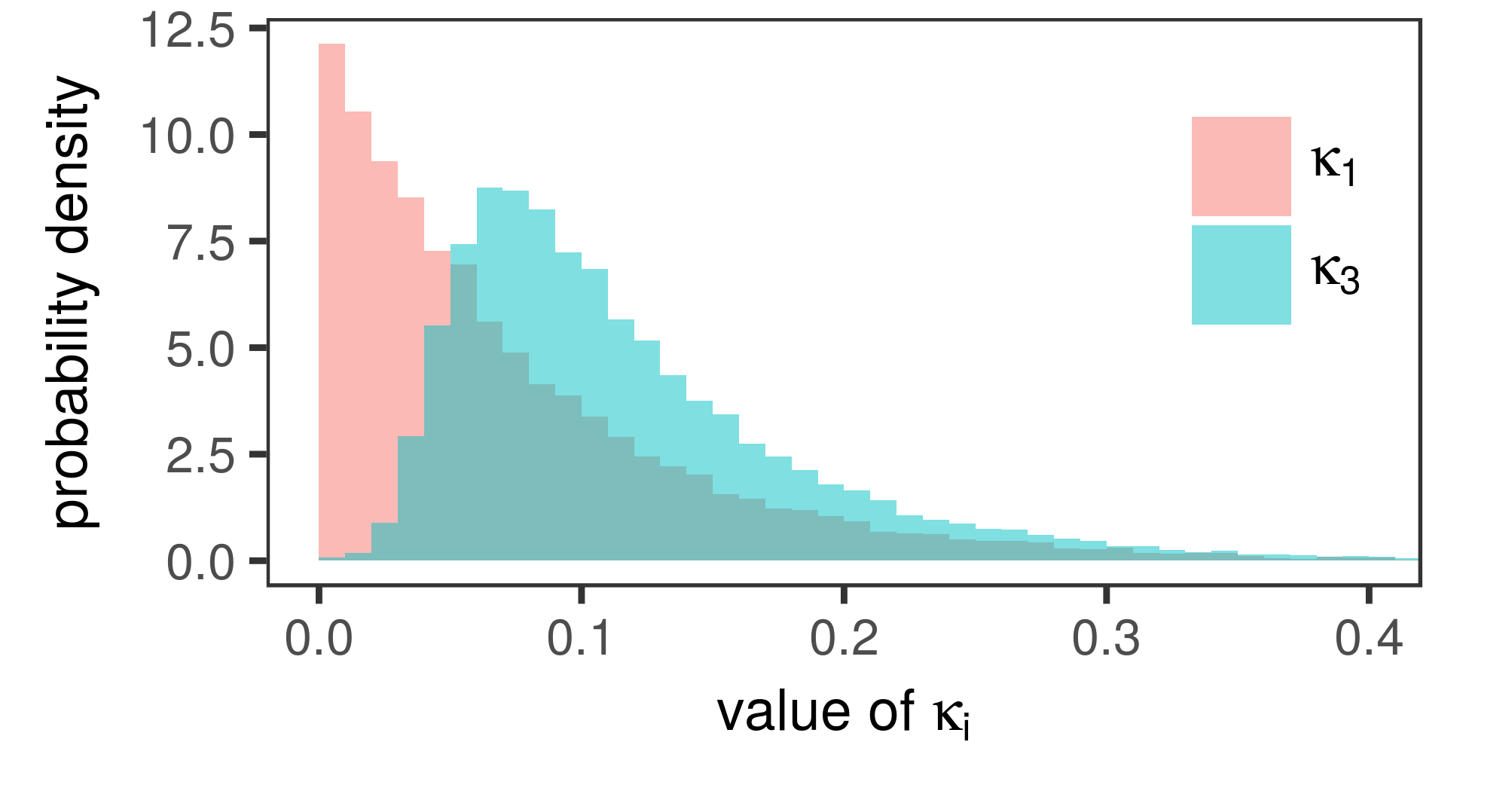}
	\caption{Estimates of the marginal posterior distributions $\rho_
	\textrm{L}(\kappavecEl{1}\,|\,\sigmaexp)$ and $\rho_\textrm{L}(\kappavecEl{3}\,|\,\sigmaexp)$ obtained from a MH chain. } 
	\label{fig:demokappahistos}
\end{figure}

Two marginal posterior distributions $\rho(\kappavecEl{i}\,|\,\sigmaexp), i\in\{1,3\}$ are shown in \cref{fig:demokappahistos}.
These distributions are considerably right-skewed.
The distribution for $\kappavecEl{3}$ rises sharply on the left and declines moderately on the right.
The sharp rise is due to the term proportional to the negated $\chi^2$-value in the second row of \cref{eq:marlike2}.
This term saturates for large enough vectors $\kappavec$.
From this point onwards, the multivariate Laplace prior $\rho_\textrm{L}(\kappavec)$ is mainly responsible for the decline.
Without the Laplace prior, the decline would be driven only by the determinant in the first row of \cref{eq:marlike2}.
As was illustrated in \cref{fig:uniformLogDens}, the rate of decline in the latter case is too low.
Please note that all these observations are specific for the assumption of a normalization uncertainty.

It may disturb to see the overall prediction running below the majority of the data points, but one should keep in mind that the method makes an assessment at the level of complete data sets.
A data set with more points does not get more weight than one with less points, hence we may call the method `democratic'.
This feature is also reasonable from an evaluation point of view.
A data set is a unit which usually comprises measurements from the same experiment.
If one data point is affected by an unrecognized systematic error, very likely the other data points are too. 
This democratic feature is also backed up by the numbers.
The solution corresponding to the overall prediction in the forth row of \cref{tab:demo_segdata} is indeed associated with the least number of non-zero normalization uncertainties.

Nevertheless, if we believe one of the data sets containing more points to be adequate, because it is more recent, comes with a detailed error analysis or measurements were performed with superior technology, we can account for this prior knowledge.
As an example, I fixed the normalization uncertainty of Bugg to $\kappavecEl{6}=0.005$ and only allowed variations of the remaining parameters in the MH algorithm.
The resulting prediction is also depicted in \cref{fig:demoovpred}.

\subsection{Beyond a Normalization Uncertainty}
\label{subsec:beyondnormalization}
Introducing a normalization uncertainty for each data set may not always be enough to correct inconsistencies.
A $\chi^2/N$ value calculated according to \cref{eq:genchisq} significantly larger than one indicates such cases.
We may use then a more general parametrization of the adjusted covariance matrix $\adjcovexp$.
We recall the assumption of vanishing correlations between data sets, so the matrix $C$ is block-diagonal with the blocks $\adjcovexpBlock{i}$.
An example of a more general parametrization of the blocks is given by
\begin{equation}
  \adjcovexpBlock{ijk}(\kappavecEl{i},\lambdavecEl{i}) = 
  \covexpBlock{i} + 
  \kappavecEl{i}^2 \exp\left( 
  	-\frac{1}{2\lambdavecEl{i}^2}  
  	(E_{ij} - E_{ik})^2
  \right) \sigma_{ij} \sigma_{ik} \,.
  \label{eq:adjcovblockGP}
\end{equation}
This form is motivated by the squared exponential function commonly used in Gaussian process regression, e.g.~\cite{rasmussen_gaussian_2006}.
The quantities $E_{ij}$ and $E_{ik}$ denote the momentum of the $j^\textrm{th}$ and $k^\textrm{th}$ data point of the experiment data set associated with $\adjcovexpBlock{i}$.
The notation for the measured cross sections $\sigma_{ij}$ and~$\sigma_{ik}$ is analogous.

The variable $\kappavecEl{i}$ denotes the relative standard deviation of the additional uncertainty component at all energies.
The length-scale $\lambdavecEl{i}$ determines how quickly the \textit{a priori} unknown error in the measured data points is allowed to change as a function of momentum.
In the limit $\lambdavecEl{i}\to\infty$, this parametrization is equivalent to the assumption of a  normalization uncertainty, compare with \cref{eq:demo_adjcovexp}.
Since the presented experiment data span a momentum range from 0.5 to 5 GeV/c, values beyond $\lambdavecEl{i}=20$ already resemble in good approximation a normalization uncertainty.
The other extreme case $\lambdavecEl{i}\to 0$ implements the assumption of white noise.
Intermediate values are well suited to capture unrecognized momentum-dependent uncertainties, such as those related to detector efficiency.

Locating the posterior maximum profits from the availability of an analytic expression for the gradient.
The computation of the gradient was discussed starting from \cref{eq:blocksofM}.
It involved the partial derivatives of the adjusted covariance matrix~$C$.
For the parametrization in \cref{eq:adjcovblockGP}, they are given by
\begin{align}
	\frac{\partial\adjcovexpBlock{i}(\kappavecEl{i},\lambdavecEl{i})}{\partial\kappavecEl{i}} &=  		
    2\kappavecEl{i} \exp\left( 
  	-\frac{1}{2\lambdavecEl{i}^2}  
  	(E_{ij} - E_{ik})^2
  \right) \sigma_{ij} \sigma_{ik} 	
\\
	\frac{\partial\adjcovexpBlock{i}(\kappavecEl{i},\lambdavecEl{i})}{\partial\lambdavecEl{i}} &=
	\frac{\kappavecEl{i}^2}{\lambdavecEl{i}^3} (E_{ij} - E_{ik})^2
	\exp\left( 
  	-\frac{1}{2\lambdavecEl{i}^2}  
  	(E_{ij} - E_{ik})^2
  \right) \sigma_{ij} \sigma_{ik}
\end{align}

The parameters $\kappavecEl{i}$ here have the same meaning as the equally named parameters linked to the magnitude of the normalization uncertainty in the previous sections.
Therefore, we can also impose the multivariate Laplace prior $\rho_\textrm{L}(\kappavec\,|\,\delta)$ in \cref{eq:demo_laplaceprior} on them.
Even though the automatic selection of $\delta$ could be done as for the normalization uncertainty, I just adopted the value $\delta_L=0.13$ for the sake of simplicity.

Some testing indicated that the marginal likelihood is rather sensitive to a length-scale $\lambdavecEl{i}$ if the points of the respective data set $\mathcal{D}_i$ are dispersed over a broad momentum range.
However, the marginal likelihood becomes insensitive if the length-scale is much larger than the momentum spread of the points.
Owing to these two observations, I opted for a multivariate Laplace prior $\rho_\textrm{L}(\lambdavec\,|\,\delta)$ with a large standard deviation $\delta=100$.
This choice ensures that the experiment data can dictate the length-scale if it matters.
And whenever the length-scale becomes too large, the prior $\rho_\textrm{L}(\lambdavec\,|\,\delta)$ regularizes the solution.

The maximization was again performed with the L-BFGS-B algorithm~\cite{byrd_limited_1995} with the same setup as described in \cref{subsec:demo_segdata}.
The parameters in $\kappavec$ were constrained to be between $10^{-4}$ and $5\times 10^{-1}$.
The parameters in $\lambdavec$ were restricted to the interval between $10^{-1}$ and $20$.
I allowed all parameters in $\kappavec$ and $\lambdavec$ to change.
Initial values for the maximization were chosen uniformly between the parameter boundaries.
I performed ten maximization attempts to ensure that a global maximum has been found.

\begin{figure}[t]
	\centering
	\includegraphics{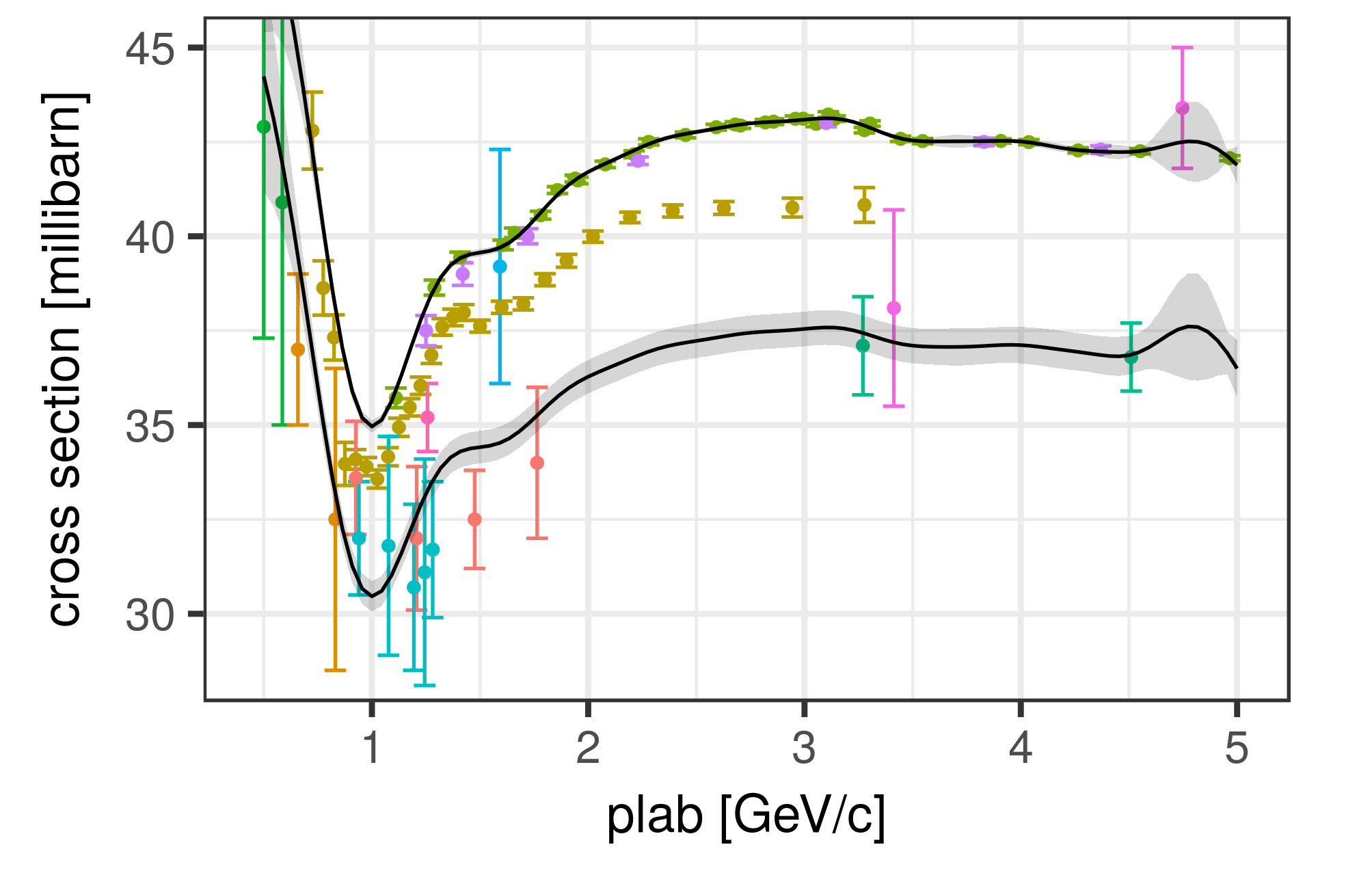}
	\caption{Lower curve shows the maximum posterior prediction and $1\sigma$-confidence band associated with the GP uncertainty in \cref{eq:adjcovblockGP} 
	The upper curve shows the maximum posterior prediction under the constraint $\lambdavecEl{6}=0.5$.} 
	\label{fig:demosegGP}
\end{figure}

The found vector $\kappavec$ was at the percent level identical to the solution in the case of normalization uncertainties, see the forth line of \cref{tab:demo_segdata}.
Also the associated prediction illustrated in \cref{fig:demosegGP} resembles that one in \cref{fig:demoovpred}.
For the data sets where $\kappavecEl{i}$ was driven towards zero, also the length-scale was driven towards the lower limit due to the influence of the prior.
Besides one exception, all other data sets with non-zero $\kappavecEl{i}$ obtained large length-scales greater than ten.
Consequently, their uncertainty parametrization resembles a normalization uncertainty.
This result is in agreement with Ockham's principle because a normalization uncertainty is already sufficient to reach consistency, see the $\chi^2/N$ values in \cref{tab:demo_segdata}, and a much simpler hypothesis than an energy-dependent uncertainty.

The exceptional case is the data set with $\kappavecEl{11}=0.08$ and $\lambdavecEl{11}=0.89$.
It contains the two pink points with large error bars on the right side of \cref{fig:demosegGP}.
Visual inspection suggests that a normalization uncertainty may not be enough to make them consistent with the overall prediction, and likely some energy-dependent error source has to be considered.
The method \textit{automatically} inferred this hypothesis by the introduction of a short length-scale.

As a final example, I applied the method another time with the constraint that $\lambdavecEl{6}=0.5$.
The upper curve in \cref{fig:demosegGP} depicts the result.
Interestingly, the fixation of the length-scale $\lambdavecEl{6}$ led to $\kappavecEl{6}\approx 0$.
The method determines that such a short-length scale is an overly complex hypothesis and therefore completely eliminates the additional uncertainty from the respective data set.
This leads in turn to the introduction of short length-scales for some of the other data sets, see \cref{tab:demo_segdataGP}.

This last section should have made clear that the method is not bound to the assumption of a normalization uncertainty.
Considering the flexibility to choose the uncertainty assumptions of the experiment data, we may regard the method better as a \textit{framework} for inference.
For instance, it would be possibly to include both a normalization uncertainty and the energy-dependent uncertainty introduced in \cref{eq:adjcovblockGP} to gain even more flexibility to adapt the uncertainty assumption of the experiments.

\begin{table}[t]
  \centering
\newcommand{\tbf}[1]{\textbf{#1}}
\begin{tabular}{llrrr}\toprule
 $i$ & reference & \# & $\kappavecEl{i}$ & $\lambdavecEl{i}$ 
\\ \midrule

 1 & SHAPIRO,PR138B,823-65 &  2 & .000 & .100 \\ 
 2 & CARVALHO,PR96,398-54 &  2 & .120 & 1.087 \\ 
 3 & DEVLIN,PRD8,136-73 & 26 & .038 & 5.762 \\ 
 4 & CHEN,PR103,211-56 &  4 & .113 & .572 \\ 
 5 & DZHELE,DOKY110,539-56 &  5 & .113 & .429 \\ 
 6 & BUGG,PR146,980-66 & 32 & .000 & .500 \\ 
 7 & ABDIVAL,NPB99,445-75 &  7 & .006 & 1.783 \\ 
 8 & KAZARINOV,JNP1,271-65 &  1 & .060 & .100 \\ 
 9 & LAW,NP9,600-59 &  1 & .000 & .100 \\ 
10 & DIDDENS,PRL9,32-62 &  2 & .100 & 8.902 \\ 
11 & PANTUEV,JNP1,134-65 &  2 & .000 & .100
\\
\bottomrule
\end{tabular}

\caption{Posterior maximum under the constraint $\lambdavecEl{6}=0.5$ using the GP uncertainty in \cref{eq:adjcovblockGP}.
The column labeled \# displays the number of points within each data set.
}   
  \label{tab:demo_segdataGP}
\end{table}

\section{Summary and Outlook}
A Bayesian method to fit models and evaluate nuclear data has been presented.
The method accounts for inconsistencies between experiment data sets by modifying their uncertainty assumptions.
The capability to correct the data has been demonstrated with an additional normalization uncertainty for each data set, and also with a more general energy-dependent uncertainty.
Due to the freedom to flexibly choose the additional uncertainty structure, we may more appropriately call the method a framework.

Related to the correction of experiment data is the possibility to segment them into consistent subsets.
Data sets whose additional uncertainty is beyond acceptable limits can be removed, so that the remaining data sets are coherent with each other.
In that respect, the multivariate Laplace prior proved to be superior over the multivariate normal prior and the uniform prior because it favors sparse solutions.

The method also allows to compute an overall prediction and the associated covariance matrix by averaging over different interpretations.
The associated uncertainties are enlarged compared to the standard GLS method, which counteracts the common problem of too small uncertainties.

It has been shown that potentially costly operations, such as the inversion of an $N\times N$ matrix with $N$ being the total number of data points, can be performed efficiently by exploiting some matrix identities.
Owing to this acceleration, the method is foreseen to work with a large corpus of data sets, and hence may be used in nuclear databases to detect inconsistencies in an automated fashion.

Future work includes the application of the method using other series expansions and physics models to better understand how the method reacts to different choices.
Because not only experiment data can be inconsistent but also models can be inaccurate, the question of model deficiencies, e.g.  \cite{neudecker_impact_2013,rochman_flatness_2014,schnabel_differential_2016}, has to be addressed, too.
To be precise, how assumptions about model deficiencies and about inconsistencies in experiment data can be best taken into account in one procedure.

Another line of research is the generalization of the method.
Besides normalization uncertainties and the energy-dependent uncertainty introduced in this paper, many other parametrizations are conceivable.
Therefore, tests with other parametrization should be performed.

Finally, with the increasing complexity of uncertainty assumptions and the increasing size of databases, possibilities for further optimization of the method likely need investigation.
This would primarily concern locating the posterior maxima and efficiently sampling from the posterior distribution.
Regarding the latter issue, adaptive sampling algorithms, such as \cite{haario_adaptive_2001}, show promise.

\appendix

\section{Acknowledgments}
This work was performed within the work package WP11
of the CHANDA project (605203) financed by the European Commission.

\section{Appendix}

\subsection{Metropolis-Hastings algorithm with symmetric proposal}
Suppose that we want to acquire a sample from the pdf $\phi(\kappavec)$.
If it is not possible to directly draw from this pdf, the MH algorithm~\cite{hastings_monte_1970} may be used.
The MH algorithm constructs a chain $\kappavec_1,\kappavec_2,\cdots, \kappavec_K$ by drawing a vector $\kappavec'$ from a proposal pdf $\psi(\kappavec'\,|\,\kappavec_i)$ on the basis of the current vector $\kappavec_i$.
In the case of a symmetric proposal distribution, i.e. $\psi(\kappavec'\,|\,\kappavec_i)=\psi(\kappavec_i\,|\,\kappavec')$, the proposed vector $\kappavec'$ is accepted with probability $\min\left(1,\phi(\kappavec')/\phi(\kappavec_i)\right)$ as the next vector $\kappavec_{i+1}$ of the chain.
Otherwise it is rejected and $\kappavec_{i}$ is taken as the next vector.
The sample represented by the chain has the pdf $\phi(\kappavec)$ as stationary distribution.


\subsection{Derivative of an inverse matrix}
The matrix $\marcov(\kappavec)$ is a function of $\kappavec$ and so is its inverse $\invmarcov(\kappavec)$.
The relation between these two matrices in terms of their components is given by
\begin{equation}
	\sum_{j} \marcov_{ij}(\kappavec) \invmarcov_{jk}(\kappavec) = \delta_{ij} \,,
\end{equation}
with $\delta_{ij}$ being one if $i=j$ and zero otherwise.
Taking the partial derivative with respect to an element $\kappa_{l}$ of $\kappavec$ gives
\begin{equation}
	\sum_{j} \left(
	\frac{\partial{\marcov_{ij}(\kappavec)}}{\partial \kappa_{l}}
	\invmarcov_{jk}(\kappavec)
	+
	\marcov_{ij}(\kappavec)
	\frac{\partial{\invmarcov_{jk}(\kappavec)}}{\partial \kappa_{l}}
	\right)  = 0 \,.
\end{equation}
This relation can be expressed in terms of matrix products,
\begin{equation}	
	\marcov(\kappavec)
	\frac{\partial{\invmarcov(\kappavec)}}{\partial \kappa_{l}}
	=
	-\frac{\partial{\marcov(\kappavec)}}{\partial \kappa_{l}}
	\invmarcov(\kappavec)
 \,.
\end{equation}
Multiplying by $\invmarcov(\kappavec)$ from the left yields
\begin{equation}
	\frac{\partial{\tilde{\mat{M}}(\kappavec)}}{\partial \kappa_{l}} =
	-\tilde{\mat{M}}(\kappavec)
	\frac{\partial{\mat{M}(\kappavec)}}{\partial \kappa_{l}}
	\tilde{\mat{M}}(\kappavec) \,.		
\end{equation}
Because $\marcov(\kappavec) = \jacob\priorcovpar\jacob^T + \adjcovexp(\kappavec)$ in this paper, we finally get
\begin{equation}
	\frac{\partial{\tilde{\mat{M}}(\kappavec)}}{\partial \kappa_{l}} =
	-\tilde{\mat{M}}(\kappavec)
	\frac{\partial{\mat{C}(\kappavec)}}{\partial \kappa_{l}}
	\tilde{\mat{M}}(\kappavec) \,.		
	\label{apx:eq:invmatderiv}
\end{equation}

\subsection{Derivative of a logarithmized determinant}

Using the chain rule, the derivate of $\ln\det\mat{M}(\kappavec)$ can be written as
\begin{equation}
	\frac{\partial \ln\det\marcov(\kappavec)}{\partial \kappavecEl{l}} =
	\frac{1}{\det\marcov(\kappavec)}
	\frac{\partial \det\marcov(\kappavec)}{\partial \kappavecEl{l}} \,.
	\label{apx:eq:logdetderive}
\end{equation}
Jacobi's formula \cite[p.~305]{harville_matrix_1997} provides us with the derivative of the determinant,
\begin{equation}
	\frac{\partial \det\marcov(\kappavec)}{\partial \kappavecEl{l}} =
	\tr \left(
	\textrm{adj}\big(\marcov(\kappavec)\big) 
	\frac{\partial\marcov(\kappavec)}{\partial\kappavecEl{l}}
	\right) \,.
	\label{apx:eq:detderive}
\end{equation}
The adjugate matrix appearing in this expression is defined by \cite[p.~192]{harville_matrix_1997}
\begin{equation}
	\marcov \, \textrm{adj}(\marcov) = \det(\marcov) \, \mathds{1}
	\;\Rightarrow\;
	\textrm{adj}(\marcov) = \det(\marcov) \mat{\marcov}^{-1}
	\label{apx:eq:adjmat}
\end{equation}
Inserting \cref{apx:eq:adjmat} into \cref{apx:eq:detderive} and the resulting expression into \cref{apx:eq:logdetderive} yields
\begin{equation}
	\frac{\partial \ln\det\marcov(\kappavec)}{\partial \kappavecEl{l}}
	= \tr \left(
	\left(\marcov(\kappavec)\right)^{-1}
	\frac{\partial\marcov(\kappavec)}{\partial\kappavecEl{l}}
	\right) \,.
\end{equation}
Because $\marcov(\kappavec) = \jacob\priorcovpar\jacob^T + \adjcovexp(\kappavec)$ in this paper, we arrive at
\begin{equation}
	\frac{\partial \ln\det\marcov(\kappavec)}{\partial \kappavecEl{l}} =
	\tr \left(
	\left(\marcov(\kappavec)\right)^{-1}
	\frac{\partial\adjcovexp(\kappavec)}{\partial\kappavecEl{l}}
	\right)	\,.
	\label{apx:eq:logdetderivative}
\end{equation}

\subsection{Matrix determinant lemma}
For the derivation of the matrix determinant lemma in the version used in this paper, note that
\begin{equation}
\begin{split}
	\det\left(\mat{A}+\mat{U}\mat{V}^T \right) &= 
	\det\left(
	\mat{A}
	\left(\mathds{1} + \mat{A}^{-1}\mat{U}\mat{V}^T \right)
	\right) \\
	&= \det(A) \det \left( \mathds{1} + \mat{A}^{-1}\mat{U}\mat{V}^T \right) \,.
\end{split}
\end{equation}
The application of Sylvester's determinant identity \cite[p.~416]{harville_matrix_1997}, i.e. $\det(\mathds{1}+\mat{A}\mat{B})=\det(\mathds{1}+\mat{B}\mat{A})$, yields
\begin{equation}
	\det\left(\mat{A}+\mat{U}\mat{V}^T \right) = \det(\mat{A})
	\det\left( \mathds{1} + \mat{V}^T \mat{A}^{-1}\mat{U} \right) \,.
\end{equation}
Now replace $\mat{U}$ by the matrix product $\mat{U}\mat{W}$ and extract $\mat{W}$ to obtain 
\begin{equation}
	\det\left(\mat{A}+\mat{U}\mat{W}\mat{V}^T \right) = \det(\mat{A})
	\det\left( \left(
	\mat{W}^{-1} + \mat{V}^T \mat{A}^{-1}\mat{U} \right)
	\mat{W} \right) \,.
\end{equation}
Making use of $\det(\mat{A}\mat{B})=\det(\mat{A})\det(\mat{B})$, we get
\begin{equation}
	\det\left(\mat{A}+\mat{U}\mat{W}\mat{V}^T \right) = 
	\det(\mat{A}) \det(\mat{W})
	\det\left(
	\mat{W}^{-1} + \mat{V}^T \mat{A}^{-1}\mat{U} \right) \,.
	\label{apx:eq:matrixdeterminantlemma}
\end{equation}

\subsection{Woobury identity}
The Woodbury identity \cite[p.~424]{harville_matrix_1997} states that
\begin{equation}
	\left(\mat{A}+\mat{U}\mat{W}\mat{V}^T\right)^{-1} =
	\mat{A}^{-1} - 
	\mat{A}^{-1} \mat{U}
	\left( 
	\mat{W}^{-1} +
	\mat{V}^T \mat{A}^{-1} \mat{U} 
	\right)^{-1}	
	\mat{V}^T \mat{A}^{-1} \,.
	\label{apx:eq:woodbury}
\end{equation}
This identity can be verified by multiplying both sides with $(\mat{A}+\mat{U}\mat{W}\mat{V}^T)$.
Applying this identity to $\marcov(\kappavec) = \jacob\priorcovpar\jacob^T + \adjcovexp(\kappavec)$ gives
\begin{equation}
	\left( \jacob\priorcovpar\jacob^T + \adjcovexp \right)^{-1} =
	\adjcovexp^{-1} - 
	\adjcovexp^{-1} \jacob 
	\left(
	\priorcovpar^{-1} + \jacob^T\adjcovexp^{-1}\jacob
	\right)^{-1}	
	\jacob^T \adjcovexp^{-1}
\end{equation}

\bibliographystyle{ans}
\bibliography{bibliography}
\end{document}